\documentclass[a4paper,USenglish]{lipics-v2021}

\usepackage[utf8]{inputenc}
\usepackage{amssymb}
\usepackage{graphicx}
\usepackage{caption}
\usepackage{enumerate}
\usepackage{relsize}
\usepackage{color}
\usepackage{verbatim}
\usepackage{amsthm}
\usepackage[ruled,linesnumbered]{algorithm2e}
\usepackage{xr}
\usepackage{tikz-cd}
\usepackage{subcaption}
\usepackage{etoolbox}
\usepackage{float}
\usepackage{xspace}
\usepackage{faktor}

\usepackage{amsmath}
\usepackage{tikz}
\usepackage{etoolbox}
\usepackage{xspace}

\DeclareMathOperator{\ind}{Ind}

\newcommand{\ie}{\emph{i.e.} }

\newcommand{\macro}[2]{ \providecommand{#1}{{\ensuremath{#2}}\xspace}}

\macro{\N}{\mathbb N}
\macro{\R}{\mathbb R}
\macro{\dB}{\bf B} 
\macro{\dG}{\bf G}
\macro{\dH}{\bf H}
\macro{\dK}{\bf K}
\macro{\dfam}{\mathcal F}
\macro{\algo}{\mathcal A}
\macro{\proc}{\odot}
\macro{\St}{\bf{St}} %Etat d'un processus dans une execution 

\newcommand{\macromath}[2]{\providecommand{#1}{{{\ensuremath{#2}}}\xspace}}
\newcommand{\rmacromath}[2]{\renewcommand{#1}{{{\ensuremath{#2}}}\xspace}}

\providecommand{\splx}[1]{\ifstrequal{#1}{1}{\sigma\xspace}%
							{\ifstrequal{#1}{2}{\tau\xspace}%
							{\ifstrequal{#1}{3}{\alpha\xspace}%
							{\ifstrequal{#1}{4}{\beta\xspace}%
							{\ifstrequal{#1}{5}{\gamma\xspace}%
							{ERROR\ :\ SIMPLEXE\ NOT\ DEFINED,\ ADD\ IT\ TO\ COMMAND}}}}}}%

\providecommand{\cplx}{C}

\providecommand{\splxMap}[1]{\ifstrequal{#1}{1}{\Phi\xspace}%
							{\ifstrequal{#1}{2}{\varphi\xspace}%
							{\ifstrequal{#1}{3}{\phi\xspace}%
							{\Phi\xspace}}}}%							

\providecommand{\In}{\mathcal{I}\xspace}
\providecommand{\Out}{\mathcal{O}\xspace}

\rmacromath{\b}{\circ}
\macromath{\n}{\bullet}
\macromath{\g}{\textcolor{gray}{\bullet}}
\macromath{\bn}{\b\!\!-\!\!\!\!\!\!-\!\!\n}
\macromath{\bg}{\b\!\!-\!\!\!\!\!\!-\!\!\g}
\macromath{\gn}{\g\!\!-\!\!\!\!\!\!-\!\!\n}
\macromath{\lnoir}{\b\!\!\rightarrow\!\!\n}
\macromath{\lblanc}{\b\!\!\leftarrow\!\!\n}
\macromath{\lall}{\b~~ \n}
\macromath{\lnall}{\b\phantom{\!\!-\!\!\!\!-\!\!}\n}
\macromath{\lok}{\b\!\!\leftrightarrow\!\!\n}
\macromath{\eps}{\varepsilon}

\macromath{\I}{\mathcal I}
\macromath{\Ou}{\mathcal O}
\macromath{\ma}{\mathcal M}
\macromath{\indlimit}{\overline{\ind}}

\usetikzlibrary{calc,
                positioning,
                patterns,
                topaths,
                arrows.meta,
                decorations.pathreplacing}

\tikzset{p/.style={circle, draw, fill=black!50,
                        inner sep=0pt, minimum width=6pt}}

\tikzstyle{proc}=[circle, draw, inner sep=0pt,
                  minimum width=8pt, line width=0.1pt]
\tikzstyle{proc1}=[circle, draw, inner sep=0pt,
                  minimum width=5pt, line width=0.1pt]
\tikzstyle{proc2}=[circle, draw, inner sep=0pt,
                  minimum width=3pt, line width=0.1pt]
\tikzstyle{proc3}=[circle, draw, inner sep=0pt,
                  minimum width=1pt, line width=0.1pt]
\tikzstyle{rel}=[>=stealth]  % style of relation "x see y"
\tikzstyle{b}=[fill=white]
\tikzstyle{g}=[fill=gray]
\tikzstyle{n}=[fill=black]

\macro{\proc}{p}
\macro{\chr}{\mathrm{Chr}}

\nolinenumbers

\title{A General Input-Dependent Colorless Computability Theorem and 
Applications to Core-Dependent Adversaries}
 \titlerunning{Input-Dependent Colorless Computability and Applications}

 \author{Yannis Coutouly}{Aix-Marseille University \and CNRS LIS UMR7020, Marseille - France }{}{}{}
 \author{Emmanuel Godard}{Aix-Marseille University \and CNRS LIS UMR7020, Marseille - France }{}{}{}

\authorrunning{Y.Coutouly,E.Godard}

\Copyright{Coutouly-Godard}
  


\EventEditors{Andrei Arusoaie, Emanuel Onica, Michael Spear, and Sara Tucci-Piergiovanni}
\EventNoEds{4}
\EventLongTitle{29th International Conference on Principles of Distributed Systems (OPODIS 2025)}
\EventShortTitle{OPODIS 2025}
\EventAcronym{OPODIS}
\EventYear{2025}
\EventDate{December 3--5, 2025}
\EventLocation{Ia\c{s}i, Romania}
\EventLogo{}
\SeriesVolume{361}
\ArticleNo{13}
\begin{document}
\maketitle

\begin{abstract}
  Distributed computing tasks can be presented with a triple
  $(\I,\Ou,\Delta)$.
  The solvability of a colorless task on the Iterated Immediate 
  Snapshot model (IIS) has been characterized by the Colorless 
  Computability Theorem \cite[Th.4.3.1]{HKRbook}. A recent 
  paper~\cite{CG-24} generalizes this theorem for any message adversaries 
  $\ma \subseteq IIS$ by geometric methods.
  
  In 2001, Mostéfaoui, Rajsbaum, Raynal, and Roy \cite{condbased} introduced 
  \emph{condition-based adversaries}. This setting 
  considers a particular adversary that will be applied only to a subset 
  of input configurations. In this setting, they studied the 
  $k$-set agreement task with condition-based $t$-resilient adversaries and 
  obtained a sufficient condition on the conditions that make $k$-Set 
  Agreement solvable.
  
  In this paper we have three contributions: 
  \begin{enumerate}
    \item We generalize the characterization of~\cite{CG-24} to 
    \emph{input-dependent} adversaries, which means that the adversaries 
    can change depending on the input configuration.
    \item We show that core-resilient adversaries of $IIS_n$ have the same 
    computability power as the core-resilient adversaries of $IIS_n$ where 
    crashes only happen at the start.
    \item Using the two previous contributions, we provide a necessary and 
    sufficient characterization of the condition-based, 
    core-dependent adversaries that can solve $k$-Set Agreement. 
  \end{enumerate}
  
  We also distinguish four settings that may appear when presenting a 
  distributed task as $(\I,\Ou,\Delta)$. Finally, in a later section, we 
  present structural properties on the carrier map $\Delta$. Such properties 
  allow simpler proof, without changing the computability power of the task. 
  Most of the proofs in this article leverage the topological framework used 
  in distributed computing by using simple geometric constructions.
\end{abstract}

\setcounter{footnote}{0}
\section{Introduction}

\subsection[short]{Topological Methods and Computability Theorems}

Since the initial work of Herlihy and Shavit~\cite{HS99}, Saks and Zaharoglou \cite{SZ}, and Borowsky and Gafni \cite{BoGa93},  showing that 
distributed computability questions are amenable to topological methods, many
important applications have been demonstrated.
This framework describes distributed problems as tasks $(\I,\Ou,\Delta)$. 
$\I$ is a colored simplicial complex of all the input 
configurations, $\Ou$ is a colored simplicial complex with all the output configurations 
and $\Delta$ is a relation that specifies, for any input, which outputs can be 
accepted. Simplicial complexes
proved to be a convenient mathematical tool to represent distributed 
situations.
In this setting, the most used model of communication is the Iterated Immediate Snapshot 
($IIS$) model, since one round of computation is simply represented by a Standard 
Chromatic Subdivision (see~\cite{HKRbook} for a detailed presentation).
One of the biggest computability results around these methods is the Asynchronous Computation 
Theorem (ACT) which says that a task $(\I,\Ou,\Delta)$ is solvable
if and only if there is a simplicial map from an iterated Standard 
Chromatic Subdivision of $\I$ to $\Ou$. 
There also exists a simpler, and more powerful, version for colorless tasks (tasks where 
the specification doesn't need the name of processes),  which states \cite[Th.~4.3.1]{HKRbook}: 
a colorless task $(\I,\Ou,\Delta)$ is solvable on the 
$IIS$ model if and only if there is a continuous map from $|\I|$ to $|\Ou|$ respecting $\Delta$.  
Such computability theorems aim to provide tight 
characterization for general tasks and allow proofs 
leveraging simplicial or topological arguments for particular tasks. 
One major line of research extends this colorless computability theorem to many
other models of communication. 

\subsection{The general message adversary line of work}

For this paper, we restrict our attention to the message adversary setting 
where the communication model has a round structure, and each round 
corresponds to a communication graph between the processes involved.
Moreover, the set of possible graphs can vary between each round, making this 
a very flexible setting. In particular, the $IIS$ model can be expressed in
this setting, as well as any subset of executions of the $IIS$ model. 
A lot of work has been done to obtain simplicial understanding of many models,
from oblivious message adversaries to core adversaries. The failure pattern has become 
more refined. For some examples see~\cite{herlihy_shavit_asynchronous_t-res_1994},
~\cite{herlihy_rajsbaum_topology_2010} or~\cite{Rieutord, KRHfairadv} 
Recently, an extension called \emph{general message adversary} has been proposed 
(see \cite{2generals-journal},~\cite{consensus-epistemo},
~\cite{CG-geoconf},~\cite{ACN23},~\cite{CG-24}). This approach can investigate 
 any subset of executions of IIS. In particular, "non-compact" sets of 
executions can be considered. A natural example is the $t$-resilient adversary : one will 
eventually get a message from $n-t$ processes, but the time 
where a process might be silent can get arbitrarily long.

The "non-compact problem" was uncovered with~\cite{2generals-journal} 
where they obtained a combinatorial and topological characterization of the 
consensus for 2-processes that observe that, for computability, some
executions seem to work in (special) pairs.
This was generalized for any 
number of processes in~\cite{CG-geoconf} with the introduction of the
\emph{geometrization} topology that interprets special pairs as a non-separable point, in the classical topological meaning.
Another direction is 
~\cite{consensus-epistemo} that provides a characterization of the solvability of Consensus 
for any number of processes for the general model of computation using an 
abstract topology on the set of executions. Then~\cite{ACN23} provides a 
colored version of an ACT-like theorem for general message adversaries,
using terminating subdivision and another ACT-like theorem for general 
models of computation, using the abstract topology. Later,~\cite{CG-24} came 
with a colorless computability theorem for general message adversaries, using 
again terminating subdivision and the \emph{geometrization} topology.

This article is built upon ~\cite{CG-24}, mostly because this approach to 
geometrization enables to consider a simple execution space that is mostly like a $\R^N$ 
space. This makes possible simple geometric reasoning as is demonstrated here. The main Theorem of 
~\cite{CG-24} is the following. Given $(\I, \Ou, \Delta)$ a colorless task, it is 
solvable on $M \subseteq IIS_n$ if and only if there is a continuous function 
$f : geo(skel^n \, I \times  \ma) \rightarrow |O|$ carried by $\Delta$, where $geo$ is the geometrization mapping. 
We can remark that, in this statement, the message adversary is independent of the input values. 
In this paper we introduce the \emph{input-dependent} setting of general 
message adversary of $IIS$ and provide a similar computability theorem for 
such model in Th.~\ref{th:independantAdv}.
An \emph{input-dependent} adversary can present executions that are different depending on the input configuration, it is a subset $\mathcal A$ of $\I\times IIS_n$. The characterization is as follows, a colorless task $(\I, \Ou, \Delta)$ is solvable under $\mathcal A$ if and only if there is a continuous function 
$f : geo(\mathcal A) \rightarrow |O|$ carried by $\Delta$, 

\subsection{Application to Open Problems}

This new computability theorem on general message adversaries is an interesting 
extension in itself, we also provide applications.
A contribution of this paper answers an open question from \cite{condbased}
about condition-based adversaries
 that were introduced  by Mostéfaoui,
  Rajsbaum, Raynal, and Roy.  This setting considers the
  solvability only for a subset of input configurations.
In \cite{condbased}, the authors investigated 
the $k$-set agreement problem within a $t$-resilient model.
They prove that if 
$C \subseteq \I$ enables to solve the task, there is a simplicial function from 
a specific complex, denoted $\mathcal{K}in(C,f,k)$, to $\Ou$. Also, two 
conditions are proposed on the 
set of inputs that make $k$-Set Agreement
solvable for a $t$-resilient model. 
Here, we propose an alternate simplicial construction 
$\mathcal{U}(\mathcal{C})$ (that is actually geometrically related to 
$\mathcal{K}in(C,f,k)$)  such that $k$-Set Agreement is solvable 
on a core-dependent model $\mathcal H$ if and only if there is a simplicial map from 
$\mathcal{U}(\mathcal{H}(\mathcal{C}))$ to $\Ou$.
Moreover, we show that knowing if a set of inputs makes
$k$-Set Agreement task solvable for a given adversary is computable.
While doing so, we show that the computability power 
of the core-resilient adversary is the same as the situation where 
crashes happen only before any communication.

\section{Models of Computation and Definitions}
\label{sec:std-defs}

\subsection{The message adversaries framework}
\label{subsec:def}
Let $n\in\N$, we consider systems with $n+1$ processes and denote 
$\Pi_n=[0,..,n]$ the set of processes. Sending a message is an asymmetric action, 
so we use directed graphs with the standard notations : let $G$, $V(G)$ is
the set of vertices, $A(G)\subset V(G)\times V(G)$ is the arcs.

\begin{definition}[Dynamic Graph and message adversary]
 Consider the collection $\mathcal G_n$ of directed graphs with vertices 
as the set $\Pi_n$.
  A \emph{dynamic graph} \dG is a sequence $G_1,G_2,\cdots,G_r,\cdots$ where
  $G_r$ is a directed graph in $\mathcal{G}_n$.
  A \emph{message adversary} $\ma$ is a set of dynamic graphs.
\end{definition}

Since $n$ will be mostly fixed through the paper, we write $\Pi$
and $\mathcal G$ when there is no ambiguity.  
We use classical vocabulary on infinite words. Let $U \subseteq \mathcal{G}$, 
$U^{*}$ is the set of finite sequences in $U$ and $U^{\omega}$ be the set of 
infinite ones. For a word $u$, $u_{|r}$ is the prefix 
of size $r$ and $u(r)$ is the $r$-th letter of the word. 
The distributed intuition behind such a graph is that $G_r$ describes whether 
there will be transmission of some messages between each pair of processes 
at the round $r$. To highlight the distributed nature of such a graph, we 
can use \emph{communication scenario} to describe a word on 
$\mathcal{G}$ and \emph{instant graph} for a letter in a word.

\subsection{Execution of a Distributed Algorithm}
\label{execution}

Given a message adversary $\ma$ and a set of initial configurations $\I$,
we define execution as an initialization step and a communication scenario.
This corresponds to a (possibly infinite) sequence of rounds of
message exchanges and corresponding local state updates.
When the initialization is clear from the context, we will use
\emph{scenario} and \emph{execution} interchangeably.

With more details, an execution of an algorithm \algo under scenario 
$w\in \ma$ and initialization $\iota\in\I$ is denoted as $\iota.w$ and 
is composed by following steps.
First, $\iota$ affects the initial state to all
processes of $\Pi$. Then the system progresses in rounds.
A round is decomposed into 3 sub-steps: sending, receiving, and updating the local state. 
 At round $r\in\N$, the processes use the \texttt{SendAll()} primitive 
 to send messages.
The fact that the corresponding receive actions, using the \texttt{Receive()}
primitive, will be successful depends on the instant graph $G$.

Let $p,q\in\Pi$. The message sent by $p$ is received by $q$ on the
condition that the arc $(p,q)\in A(G)$.  Then, all processes update
their state according to the received values and \algo.  Note that it
is assumed that $p$ always receives its own value, which is $(p,p)\in
A(G)$ for all $p$ and $G$. However, this might be
implicit for clarity and brevity. 
We denote the local state of a process $p$ in an execution $w = \iota.u$
 at the round $r$ of the algorithm as $\mathbf s_p(\iota.u[r])$.
$\mathbf s_p(\iota.\varepsilon) = \iota(p)$ represents the initial 
state of $p$ in $\iota$, where $\varepsilon$ is the empty word. 
Note that in our framework, processes have identities (``colors'') and
can therefore distinguish identical values sent by different
processes. This model is denoted as the colored model of computation.

\subsection{Relevant communication model}

The message adversary framework allows us to describe a wide array of 
communication models. Like many other distributed computing papers that 
adopt a topological approach, the central model in this one is the Iterated 
Immediate Snapshot ($IIS$) model. It was first introduced as a
(shared) memory model, which has been proved equivalent to the
message adversary below first as tournaments and iterated tournaments
~\cite{BGequivIIS,messadv}, then as this message adversary
~\cite{HKRbook,DCcolumn}.  See also~\cite{Rajsbaum-iterated} for a survey of 
the reductions involved in these layered models.
Given a graph $G$, we denote by
$In_G(a) = \{ b\in V(G) \mid (b,a)\in A(G)\}$ the set of incoming vertices of 
$a$ in $V(G)$. A graph $G$ has the \emph{containment Property} if for all 
$a,b\in V(G)$, $In_G(a)\subset In_G(b)$ or $In_G(b)\subset In_G(a)$.
We say that a graph $G$ has the \emph{Immediacy Property} if for all 
$a,b,c\in V(G)$, $(a,b), (b,c)\in A(G)$ implies that $(a,c)\in A(G)$.

\begin{definition}[IIS model~\cite{HKRbook}]
  We set
  $ImS_n = \{G\in\mathcal G_n\mid G$ has the Immediacy and
 \mbox{Containment} properties~$\}$.
The Iterated Immediate Snapshot message adversary for $n+1$ processes is the
message adversary $IIS_n=ImS_n^\omega$.
\end{definition}

The setting of this paper is the general sub-message adversaries of
the $IIS$ model, which is $\ma \subseteq IIS_n$. Many studied adversaries in the literature
can be represented as such, like \emph{oblivious adversary},
\emph{t-resilient adversary} or \emph{core-resilient adversary} or any
'situation-specific' adversary. We also use the
terminology \emph{crash} (which is, strictly speaking, irrelevant for
messages adversaries) for a process that is not heard of by all other
processes for an infinite amount of time.

As an example with two
processes $\b$ and $\n$, we can define the message adversary $IIS_1 =
\{\lok,\lblanc,\lnoir\}^\omega$. In the execution
$\lok\lblanc^\omega$, process \b is considered crashed starting from the second
round. In the execution $\lok(\lblanc\lnoir)^\omega$, no process is
crashed.  A message adversary like
$\ma_1=\{\lok^\omega\}\cup\{\lok\}^*(\{\lblanc^\omega,
\lnoir^\omega\})$, that represent a system with two synchronized processes, where at
most one of the processes may crash is a strict sub-adversary of $IIS_1$
since $\ma_1\subsetneq IIS_1$. \\
An iterated $t$-resilient adversary is a set of executions where at
most $t$ processes may ``crash'' during the execution. \\
Let $Q(w)$ the set of processes that are seen by all processes an infinite 
number of times in $w$

\begin{definition}[Iterated $t$-resilient adversary]
Let $R_t^n= \{ w \in IIS_n \, | \, \exists Q(w) \in \Pi$ such that $\#Q(w)
\geq n+1-t\}$ be the $t$-resilient model on $IIS_n$.
\end{definition}

Core-resilient adversaries are a generalization of $t$-resilient adversaries. 

\begin{definition}[Core-resilient adversary]
Let $P$ an inclusion-closed collection of sets of processes, a core-resilient 
adversary on $P$ is the following set of executions 
$\mathcal{H}_P = \{ w \in IIS_n \, | \, \Pi_n\setminus Q(w)\in P\}$. 
\end{definition}

\section{Abstract simplicial complexes and colorless tasks}
\label{sec:abstractSimplices}

We start by restating some standard definitions of combinatorial topology.

\begin{definition}[Abstract simplicial complex]\cite[Def 3.2.1]{HKRbook}
Let $V$ be a set, and $\cplx$ a collection of finite subsets of $V$.
$C$ is an abstract simplicial complex on $V$ if
%\begin{enumerate}
%\item
  $\forall\splx1 \in C, \forall \splx2 \subseteq \splx1,$ we have $\splx2 \in C;$
%	\item
and $\forall v \in V, \{v\} \in C.$
%\end{enumerate}
\end{definition}

An element of $V$ is a \emph{vertex} of $\cplx$ and $V(\cplx)$ denotes the set of vertices of $\cplx$. A set $\splx1 \in \cplx$ is a \emph{simplex} where $dim\ \splx1 $ is the number of vertices in $\splx1$ minus one. We say that $\splx1$ is a \emph{facet} if there is no other simplex that contains $\splx1$. 
If $\cplx_1 \subseteq \cplx_2$ then we say that $\cplx_1$ is a 
\emph{subcomplex} of $\cplx_2$, a complex is \emph{pure} if all facets 
have the same dimension.

\begin{definition}[Simplicial map]\cite[Def 3.2.2]{HKRbook}
Let $\cplx_1,\cplx_2$ be two simplicial complexes, a simplicial map is a map $\splxMap1 : V(\cplx_1) \rightarrow V(\cplx_2)$ such that $\forall \splx1 \in \cplx_1, \splxMap1(\splx1) \in \cplx_2.$
\end{definition}

\begin{definition}[Carrier map]\cite[Def 3.4.1]{HKRbook}
  Let $\cplx_1,\cplx_2$ be two simplicial complexes, a carrier map $\splxMap1 : \cplx_1 \rightarrow 2^{\cplx_2}$
  associates each simplex to a subcomplex of $\cplx_2$.
  The map is monotone,
$\forall \splx1, \splx2 \in \cplx_1,$  $\splx1 \subseteq \splx2$ implies $\splxMap1(\splx1) \subseteq \splxMap1(\splx2)$.
\end{definition} 

The pair $(\cplx_1,\chi_{\cplx_1})$ is a \emph{chromatic complex} if
$\cplx_1$ is a complex and the function $\chi_{\cplx_1} : V(\cplx_1)
\rightarrow \Pi$ has the property that $\forall \splx1 \in \cplx_1,
\forall v_1,v_2 \in V(\splx1), v_1 \neq v_2 \Leftrightarrow
\chi_{\cplx_1}(v_1) \neq \chi_{\cplx_1}(v_2)$.

%The pair $(\cplx_1,\chi_{\cplx_1})$ is a chromatic complex.

The \emph{border} of a simplex $\splx1$, is $\partial(\splx1) = \{ \splx2 \in \splx1 | dim(\splx2) = \dim(\splx1) -1 \}$. 
A \emph{$\ell$-skeleton} of $\cplx_1$ is the collection of the simplices of dimensions equal or less than $\ell$, we write $skel^{\ell}(\cplx_1)$.
The \emph{star} of a simplex $\splx1 \in \cplx_1$ is $St(\splx1,\cplx_1) = 
\bigcup_{\splx2 \in \cplx_1, \splx1 \subseteq \splx2} \splx2 $. 
The \emph{Link} of a simplex $\splx1$ is 
$Lk(\splx1,\cplx1) = \{ \splx2 \in St(\splx1, \cplx1) \, | \, 
\splx1 \cap \splx1 = \emptyset\}$. 
A simplicial map $\varphi : \cplx_1 \rightarrow \cplx_2$ is \emph{carried} by 
$\splxMap1$ if $\forall \splx1 \in \cplx_1, \varphi(\splx1) 
\in \splxMap1(\splx1)$.

This simplicial framework is used to describe \emph{distributed task}, which 
are distributed problems. Let $V_{in}$ be the domain of input values and 
$V_{out}$ be the domain of output values. Here we focus on colorless tasks.

\begin{definition}[Colorless Task]\cite[Def 4.2.1]{HKRbook} 
A \emph{colorless task} is a triple $(\mathcal{I},\mathcal{O},\Delta)$ where : 
\begin{itemize}
    \item $\mathcal{I}$ is a simplicial complex, with vertices 
    $V_{in}$,
    \item $\mathcal{O}$ is a simplicial complex, with vertices $V_{out}$,
    \item $\Delta : \In \rightarrow 2^{\mathcal{O}}$ is a carrier map.
\end{itemize}
\end{definition}

The complex $\I$ is called the \emph{input complex}, the complex $\Ou$
the \emph{out complex} and $\Delta$ encodes the specification of the
task. Colorless tasks correspond to a large family of problems where
the number of occurrences of a value is not to take into account. The
standard $k-$set agreement problem is a colorless task (only the number of
different values is constrained). The renaming task is not a colorless
task (each name should be unique in the output).
  
\section{Colorless tasks into a colored world}
\label{sec:colorlessToColored}

This section explicit the multiple settings that are used in distributed 
computing with combinatorial topology in particular in the context of 
colorless tasks.% $(\I,\Ou,\Delta)$.

%% Processes may have no 
%% identifier, we call this the unique-value setting. Or the processes have an 
%% identifier, we are in the multi-value setting. \\ 
%% In the second subsection, we define 4 classes of adversaries from a colorless 
%% task: The \emph{standard setting} $\ma_1 = \I \times IIS$, the \emph{submodel 
%% setting} $\ma_2 = \I \times \ma_3$ with $\ma_3 \subseteq IIS$. The 
%% \emph{condition-based} setting $\ma_4 = C \times \ma_3$ with $C \subseteq \I$ 
%% and the \emph{input-dependent setting} $\ma_4 \subseteq \I \times IIS$.

\subsection{Encoding colorless tasks with processes}
Even though we consider only colorless tasks, our model of computation
is colored (see Section~\ref{sec:std-defs}). In the colored
computation model, the input configurations of a given colorless task
are usually all the possible assignments of the initial values of the
processes, so an initial value could be assigned many times.  This means that the colored initial complex is not exactly \I.
Since we
will consider the very general input-dependent model, and a colored model of computation,
we need a clear representation of the state of each process.

%% In this
%% section, we present the two ways to encode an input from a colorless
%% task into a colored setting.

Consider a colorless task $(\I,\Ou,\Delta)$, and set $V_{in} = V(\I)$ and 
$V_{out}= V(\Ou)$. We assume wlog that there exists a total order $\prec$
on $V_{in}$ and that we can describe $V_{in}$ as $\{v_0,v_1,\cdots,v_n\}$, where
$v_i\prec v_j$ when $i<j$.  
The first setting is called \emph{unique-value}.
From a set of processes $\Pi = [0,n]$, there is only one process that can 
have a given initial value, and each process has exactly one value initially.
This yields a colored simplicial complex $\mathcal{V}(\I)$ with vertices
$\{(i,v_i)\mid i\in [n]\}$ and simplices $((i_0,v_{i_0}),\dots,(i_d,v_{i_d}))$ 
whenever $\{v_{i_0},\dots,v_{i_d}\}$ is a simplex of $\I$. This colored simplicial complex is actually \I, adding colours in the enumeration order $\prec$.
The second setting is called \emph{multi-value}, which allows
different processes to have the same initial value, including in 
the same initial configuration. This forms a pseudosphere of $\I$, which 
is the following colored simplicial complex $\mathcal P_n(\I)$. The set of vertices is 
$\Pi \times V(\I)$. A set
$\{(i_0,v_{i_0}),\dots,(i_d,v_{i_d})\}$, with $i_j<i_{j'}$ for $j<j'$, is a simplex of $\mathcal P_n(\I)$ whenever
$\{v_{i_0},\dots,v_{i_d}\}$ is a simplex of $\I$. Note that here
$\{(i_0,v_{i_0}),\dots,(i_d,v_{i_d})\}$ is a simplex of dimension $d$
whereas $\{v_{i_0},\dots,v_{i_d}\}$ could be a simplex of dimensions
less than $d$.
Fig.~\ref{fig:subModelExample} illustrates the Binary Consensus 
task as encoded by a relation between input and output complexes.

Fig.~\ref{fig:inProc_vs_inVal} gives two representations of the 
encoding of Binary Consensus in the colored model of computation (with \b as color $0$ and \n as color $1$),
one in the unique value setting, and the other one 
in the multi-value setting.

\begin{figure}[ht]
\centering
\begin{tikzpicture}[scale=.35]
%%%%%%%%%%%%%%%%%%%%%%%%%%%%%% Complex I
      \node[] at (-1,2.5) {$\mathcal{I}$};

      \node[proc2,b,label=left:$\{0\}$] (1) at (0, 5.0) {};
      \node[proc2,b,label=left:$\{1\}$] (2) at (0, 0) {};
      \draw (1) -- (2);

%%%%%%%%%%%%%%%%%%%%%%%%%%%%%%%% Complex O
      \node[] at (8.5,2.5) {$\mathcal{O}$};
      \node[proc2,b,label=right:$\{1\}$] (4) at (7, 0) {};
      \node[proc2,b,label=right:$\{0\}$] (5) at (7, 5.0) {};
      %\draw (4) -- (6);
      %\draw (5) -- (7);
      
%%%%%%%%%%%%%%%%%%%%%%%%%%%%%%%% Carrier map

      \node[blue] at (3,4) {$\Delta$};
      
      \node[] (12) at (0,2.5) {};
      \node[] (13) at (5.5,2.5) {};      
      \draw[blue,label=north:${\Delta}$] (12) -- (13);
      
      \node[] (14) at (5,2.5) {};
      \node[] (15) at (5,2) {};
      \node[] (16) at (5,3) {};
      \draw[->,blue] (16) to[bend right] (4);
      \draw[->,blue] (15) to[bend left] (5);
      \draw[->,blue] (1) -- (5);
      \draw[->,blue] (2) -- (4);
\end{tikzpicture}
\caption{The Binary Consensus task.}
\label{fig:subModelExample}
\end{figure}
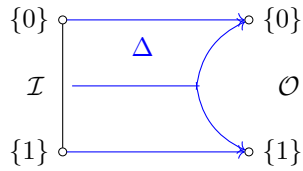

\begin{figure}[ht]
\centering
\begin{subfigure}[t]{.49\textwidth}
  \begin{tikzpicture}[scale=.35]
    %%%%%%%%%%%%%%%%%%%%%%%%%%%%%% Complex I
          \node[] at (-1,2.5) {$\mathcal V(\I)$};
          \node[proc,b,label=left:$\{0\}$] (1) at (0, 5.0) {};
          \node[proc,n,label=left:$\{1\}$] (2) at (0, 0) {};
          \draw (1) -- (2);

%          \draw (4.5,-1) -- (4.5,6.0);
    %%%%%%%%%%%%%%%%%%%%%%%%%%%%%%%% Complex O
          \node[] at (8.5,2.5) {$\mathcal{O}$};
          \node[proc2,b,label=right:$\{1\}$] (4) at (7, 0) {};
          \node[proc2,b,label=right:$\{0\}$] (5) at (7, 5.0) {};
          %\draw (4) -- (6);
          %\draw (5) -- (7);
          
    %%%%%%%%%%%%%%%%%%%%%%%%%%%%%%%% Carrier map
        \node[blue] at (3,4) {$\Delta$};

          \node[] (12) at (0,2.5) {};
          \node[] (13) at (5.5,2.5) {};      
          \draw[blue] (12) -- (13);
          
          \node[] (14) at (5,2.5) {};
          \node[] (15) at (5,2) {};
          \node[] (16) at (5,3) {};
          \draw[->,blue] (16) to[bend right] (4);
          \draw[->,blue] (15) to[bend left] (5);
          \draw[->,blue] (1) -- (5);
          \draw[->,blue] (2) -- (4);
    \end{tikzpicture}
    \caption{Binary Consensus in the unique value setting}
  \end{subfigure}
\begin{subfigure}[t]{0.49\linewidth}
  \begin{tikzpicture}[scale=.3]
    %%%%%%%%%%%%%%%%%%%%%%%%%%%%%% Complex I
        \node[] at (2.5,1) {$\mathcal{P}(\mathcal{I}$)};
      \node[proc,n,label=left:$\{1\}$] (0) at (0, 0) {};
          \node[proc,b,label=left:$\{0\}$] (1) at (0, 5.0) {};
          \node[proc,b,label=right:$\{1\}$] (2) at (5.0, 0) {};
          \node[proc,n,label=right:$\{0\}$] (3) at (5.0, 5.0) {};
          \draw (0) -- (2);
          \draw (1) -- (0);
          \draw (3) -- (2);
          \draw (1) -- (3);

 %         \draw (8.5,-1) -- (8.5,6.0);
      %%%%%%%%%%%%%%%%%%%%%%%%%%%%%%%% Complex O
      \node[] at (15,2.5) {$\mathcal{O}$};
      \node[proc2,b,label=south:$\{1\}$] (4) at (15, 0) {};
      \node[proc2,b,label=north:$\{0\}$] (5) at (15, 5.0) {};
          
    %%%%%%%%%%%%%%%%%%%%%%%%%%%%%%%% Carrier map
          \node[blue] at (9.5,4) {$\Delta$};

          \node[] (8) at (2.5,5) {};
          \node[] (9) at (14.5,5) {};
          \draw[->,blue] (8) to[bend left] (9);
          
          \node[] (10) at (2.5,0) {};
          \node[] (11) at (14.5,0) {};
          \draw[->,blue] (10) to[bend right] (11);
          
          \node[] (12) at (0,2.5) {};
          \node[] (13) at (9,2.5) {};      
          \draw[blue] (12) -- (13);
          
          \node[] (14) at (8,2.5) {};
          \node[] (15) at (8,2) {};
          \draw[->,blue] (14) to[bend right] (9);
          \draw[->,blue] (14) to[bend left] (11);
    \end{tikzpicture}
    \caption{Binary Consensus in the multi-value setting}
  \end{subfigure}
  \caption{\label{fig:inProc_vs_inVal}}
\end{figure}

Moving from one setting to another can be done using the function 
$GIV : \I_P \rightarrow \I_V$ (as Get Input Value) to associate a colored 
simplex to a colorless simplex that contains the same input values :
$GIV(\{(p_0,v_0), (p_1,v_1), \dots , (p_n,v_n)\}) = \{ v_0, v_1, \dots
v_n\}$. Moreover, the carrier map for the multi-value setting $\Delta_p$ needs 
some adjustments, $\forall \splx1 \in \mathcal{P}_n(\I), 
\Delta_p(\splx1) = \Delta(GIV(\splx1))$. Which makes sense, in a colorless task 
the value of a simplex decides its possible output.

%% An input complex where the vertices correspond to an input value and a process 
%% will be named as \emph{input-process} complex. From an input value complex $\I_v$, 
%% we want to represent every combination of (input value, process name) 
%% possibles: Let $
%% \I_p = \mathcal{P}(\I_v) = \{ \{(p_0,v_0), (p_1,v_1), \dots , (p_n,v_n)\}
%% \, | \, \forall i,j \in [0,n], p_i,p_j \in \Pi$ and $p_i \neq p_j$ moreover $ 
%% \{ v_0, v_1 \dots v_n\} \in \I_v \}$. This operation is known as the doing 
%% the pseudo-sphere of the simplicial complex $\I_v$. We stress that 
%% $\{ v_0, v_1 \dots v_n\}$ this simplex can contain identical value and hence 
%% be of smaller dimension than $n$. 
%% We use the function $GIV : \I_p \rightarrow \I_v$ (as Get Input Value)
%% to associate all the simplex that contain the same input value : 
%% $GIV(\{(p_0,v_0), (p_1,v_1), \dots , (p_n,v_n)\}) = 
%% \{ v_0, v_1, \dots v_n\}$. With this relation we can obtain 
%% an input-value complex from an input-process complex : 
%% Let $\zeta(\I_p) = \{ GIV(\splx1) \, | \, \forall \splx1 \in \I_p, \}$ 
%% Also, from a colorless task $(\I_v,\Ou,\Delta_v)$ we can define the 
%% corresponding task with an input-process complex : $(\I_p,\Ou,\Delta_p)$
%% With $\I_p = \mathcal{P}(\I_v)$ and $\forall \splx1 \in \I_p, 
%% \Delta_p(\splx1) = \Delta_v(GIV(\splx1))$. We denote this relation between task
%% by saying that $(I_p,\Ou,\Delta_p)$ is the \emph{colored-equivalent} of 
%% $(\I_v,\Ou,\Delta)$, in the over direction we say that $(I_v, \Ou, \Delta_v)$
%% is the \emph{colorless-equivalent} of $(I_p,\Ou,\Delta_p)$. 

\subsection{Problem Statements}

This paper considers 4 classes of adversaries that can be applied to a 
colorless task $(\I,\Ou,\Delta)$. The \emph{standard setting} has a 
message adversary $\ma_1 = IIS_n$ while the \emph{sub-model setting} 
has a message adversary $\ma_1 \subseteq IIS$. Both can apply to a unique 
value context ($\ma = \mathcal{V}(\I) \times \ma_1)$ or a multi-value context 
($\ma = \mathcal{P}_n(\I) \times \ma_1$).
These two contexts are considered interchangeable in 
the literature, since in the standard setting, they are simply equivalent, 
as we will prove on 
Prop.~\ref{prop/equivum}. In the input-dependent submodel setting,
 one has to be more careful.
 The results of~\cite{CG-24} were presented in the unique-value setting.
The \emph{condition-based setting} has a message adversary $\ma_1 
\subseteq IIS_n$ and $C \subseteq \mathcal{P}_n(\I)$ to form executions on 
$\ma = C \times \ma_1$. This corresponds to adding a condition of distribution of 
initial values that are not valid. This is the setting of~\cite{condbased} 
with $\ma_1 = \mathcal{R}^t_n$ (the $t$-resilient model).
Finally, the \emph{input-dependent setting} considers execution in 
$\ma \subseteq \I \times IIS_n$, which adds the possibility of changing the 
message adversary depending on the input configuration. This setting 
encompasses all previously described settings. 

\begin{figure}[t]
  \centering
\begin{subfigure}[t]{0.24\linewidth}
  \begin{tikzpicture}[scale=.3]
    %%%%%%%%%%%%%%%%%%%%%%%%%%%%%% Complex I
        \node[] at (2.5,3.5) {$\mathcal{C}_1$};
      \node[proc,n,label=left:$\{1\}$] (0) at (0, 0) {};
          \node[proc,b,label=left:$\{0\}$] (1) at (0, 5.0) {};
          \node[proc,b,label=right:$\{1\}$] (2) at (5.0, 0) {};
          \node[proc,n,label=right:$\{0\}$] (3) at (5.0, 5.0) {};
          \draw (0) -- (2);
          \draw (1) -- (3);

    %%   %%%%%%%%%%%%%%%%%%%%%%%%%%%%%%%% Complex O
    %%   \node[] at (15,2.5) {$\mathcal{O}$};
    %%   \node[proc2,b,label=south:$\{1\}$] (4) at (15, 0) {};
    %%   \node[proc2,b,label=north:$\{0\}$] (5) at (15, 5.0) {};
          
    %% %%%%%%%%%%%%%%%%%%%%%%%%%%%%%%%% Carrier map
    %%     \node[blue] at (10,5) {$\Delta$};

    %%       \node[] (8) at (2.5,5) {};
    %%       \node[] (9) at (14.5,5) {};
    %%       \draw[->,blue] (8) to[bend left] (9);
          
    %%       \node[] (10) at (2.5,0) {};
    %%       \node[] (11) at (14.5,0) {};
    %%       \draw[->,blue] (10) to[bend right] (11);
          
    \end{tikzpicture}
    \caption{Condition $\mathcal{C}_1$\label{cond1}}
  \end{subfigure}
\begin{subfigure}[t]{0.24\linewidth}
\begin{tikzpicture}[scale=.3]
  %%%%%%%%%%%%%%%%%%%%%%%%%%%%%% Complex I
      \node[] at (2.5,3.5) {$\mathcal{C}_2$};
        \node[proc,b,label=left:$\{0\}$] (1) at (0, 5.0) {};
        \node[proc,b,label=right:$\{1\}$] (2) at (5.0, 0) {};
        \node[proc,n,label=right:$\{0\}$] (3) at (5.0, 5.0) {};
        \draw (3) -- (2);
        \draw (1) -- (3);

  %%   %%%%%%%%%%%%%%%%%%%%%%%%%%%%%%%% Complex O
  %%   \node[] at (15,2.5) {$\mathcal{O}$};
  %%   \node[proc2,b,label=south:$\{1\}$] (4) at (15, 0) {};
  %%   \node[proc2,b,label=north:$\{0\}$] (5) at (15, 5.0) {};
        
  %% %%%%%%%%%%%%%%%%%%%%%%%%%%%%%%%% Carrier map
  %%       \node[blue] at (9.5,4) {$\Delta$};

  %%       \node[] (8) at (2.5,5) {};
  %%       \node[] (9) at (14.5,5) {};
  %%       \draw[->,blue] (8) to[bend left] (9);

  %%       \node[] (10) at (5,0) {};
  %%       \node[] (11) at (14.5,0) {};
  %%       \draw[->,blue] (10) to[bend right] (11);
        
  %%       \node[] (12) at (5,2.5) {};
  %%       \node[] (13) at (9,2.5) {};      
  %%       \draw[blue] (12) -- (13);
        
  %%       \node[] (14) at (8,2.5) {};
  %%       \node[] (15) at (8,2) {};
  %%       \draw[->,blue] (14) to[bend right] (9);
  %%       \draw[->,blue] (14) to[bend left] (11);
  \end{tikzpicture}
  \caption{Condition $\mathcal{C}_2$\label{cond2}}
\end{subfigure}
\begin{subfigure}[t]{0.24\linewidth}
\begin{tikzpicture}[scale=.3]
  %%%%%%%%%%%%%%%%%%%%%%%%%%%%%% Complex I
      \node[] at (1.35,3.5) {$geo(\mathcal{C}_2\times \ma_1)$};
        \node[proc,b,label=left:$\{0\}$] (1) at (0, 5.0) {};
        \node[proc,b,label=right:$\{1\}$] (2) at (5.0, 0) {};
        \node[proc,n,label=right:$\{0\}$] (3) at (5.0, 5.0) {};
        \draw (3) -- (2);
        \draw (1) -- (3);

        \node[red] at (2.5,5) {$\times$};        
        \node[red] at (5,2.5) {$\times$};                
  
  %%   %%%%%%%%%%%%%%%%%%%%%%%%%%%%%%%% Complex O
  %%   \node[] at (15,2.5) {$\mathcal{O}$};
  %%   \node[proc2,b,label=south:$\{1\}$] (4) at (15, 0) {};
  %%   \node[proc2,b,label=north:$\{0\}$] (5) at (15, 5.0) {};
        
  %% %%%%%%%%%%%%%%%%%%%%%%%%%%%%%%%% Carrier map
  %%       \node[blue] at (9.5,4) {$\Delta$};

  %%       \node[] (8) at (2.5,5) {};
  %%       \node[] (9) at (14.5,5) {};
  %%       \draw[->,blue] (8) to[bend left] (9);

  %%       \node[] (10) at (5,0) {};
  %%       \node[] (11) at (14.5,0) {};
  %%       \draw[->,blue] (10) to[bend right] (11);
        
  %%       \node[] (12) at (5,2.5) {};
  %%       \node[] (13) at (9,2.5) {};      
  %%       \draw[blue] (12) -- (13);
        
  %%       \node[] (14) at (8,2.5) {};
  %%       \node[] (15) at (8,2) {};
  %%       \draw[->,blue] (14) to[bend right] (9);
  %%       \draw[->,blue] (14) to[bend left] (11);
  \end{tikzpicture}
  \caption{Geometrization in a condition-based setting.\label{gencond}}
\end{subfigure}
\begin{subfigure}[t]{0.24\linewidth}
\begin{tikzpicture}[scale=.3]
  %%%%%%%%%%%%%%%%%%%%%%%%%%%%%% Complex I
      \node[] at (2.5,3.5) {$geo(\ma_2)$};
        \node[proc,b,label=left:$\{0\}$] (1) at (0, 5.0) {};
        \node[proc,b,label=right:$\{1\}$] (2) at (5.0, 0) {};
        \node[proc,n,label=right:$\{0\}$] (3) at (5.0, 5.0) {};
        \draw (3) -- (2);
        \draw (1) -- (3);
  
        \node[red] at (5,2.5) {$\times$};                
  
  %%   %%%%%%%%%%%%%%%%%%%%%%%%%%%%%%%% Complex O
  %%   \node[] at (15,2.5) {$\mathcal{O}$};
  %%   \node[proc2,b,label=south:$\{1\}$] (4) at (15, 0) {};
  %%   \node[proc2,b,label=north:$\{0\}$] (5) at (15, 5.0) {};
        
  %% %%%%%%%%%%%%%%%%%%%%%%%%%%%%%%%% Carrier map
  %%       \node[blue] at (9.5,4) {$\Delta$};

  %%       \node[] (8) at (2.5,5) {};
  %%       \node[] (9) at (14.5,5) {};
  %%       \draw[->,blue] (8) to[bend left] (9);

  %%       \node[] (10) at (5,0) {};
  %%       \node[] (11) at (14.5,0) {};
  %%       \draw[->,blue] (10) to[bend right] (11);
        
  %%       \node[] (12) at (5,2.5) {};
  %%       \node[] (13) at (9,2.5) {};      
  %%       \draw[blue] (12) -- (13);
        
  %%       \node[] (14) at (8,2.5) {};
  %%       \node[] (15) at (8,2) {};
  %%       \draw[->,blue] (14) to[bend right] (9);
  %%       \draw[->,blue] (14) to[bend left] (11);
\end{tikzpicture}
  \caption{Geometrization in the input-dependent setting \label{indep}}
\end{subfigure}
\caption{Examples of the different possible settings.
\label{fig:conditionBasedExample}}
\end{figure}
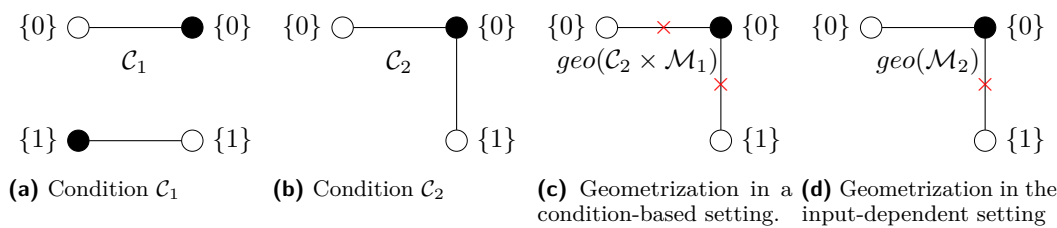

In Figure \ref{fig:conditionBasedExample}, we present examples of our
various setting for the Binary Consensus. In Fig.~\ref{cond1}, for
condition $\mathcal{C}_1$, we remove the possibility of $\n$ and $\b$
to have different initial values. On Fig.~\ref{cond2}, for condition
$\mathcal{C}_2$, we remove all input configurations where $\n$ starts
with the value $1$.  We denote $\ma_1 = IIS_1\setminus
\{\lok^\omega\}$. Fig.~\ref{gencond} presents the geometrization of sub-model 
condition-based adversaries $\mathcal C_2\times \ma_1$.
Intuitively, the red cross represents the missing execution $\lok^\omega$, see
later in Section~\ref{sec:geo}.
We define $\ma_2$ to be the input dependent
model where the possible inputs are from $\mathcal C_2$, and the
possible executions are $IIS_1$ when the two initial values are $0$,
and where the possible executions are $\ma_1$ when the two initial
values are different. The geometrization of $\ma_2$ is given in 
Fig.~\ref{indep}.

 %Directly, it appears that the Binary consensus is 
 %solvable on the condition $\mathcal{C}_1$ but not in the condition 
 %$\mathcal{C}_2$.

\begin{definition}[Solvability of a Colorless Task]
Given a colorless task $(\I,\Out,\Delta)$, it is solvable in the
\emph{input-dependent setting} against executions 
$\ma \subseteq \I \times IIS_n$ if there is a colored algorithm \algo such
 that for any execution $\iota.w\in \ma$, there exists $u$ a prefix of $w$ 
 such that the state of the system $\{\mathbf s_0(\iota.u),\dots,
 \mathbf s_n(\iota.u)\} = out$ satisfies the specification of the task, 
 i.e. $out\in\Delta(GIV(\iota))$.
\end{definition}

%\subsection{Equivalence of Settings}

The unique-value setting and the multi-value setting are
equivalent on IIS from a computability point of view.

%This is also the
%case when the adversary is view-stable (defined later).

\begin{proposition}\label{prop/equivum}
  There is an algorithm to solve $(\I,\Out,\Delta)$ on $IIS_n$ with
  condition $\mathcal V(\I)$ if and only if there is an algorithm to solve
  $(\I,\Out,\Delta)$ on $IIS_n$ with condition $\mathcal P_n(\I)$.
\end{proposition}
\begin{proof}
  Since $\mathcal V(\I)\subset\mathcal P_n(\I)$, one direction is
  obvious. Consider that we have an algorithm $Algo$ with input $\mathcal
  V(\I)$, then we extend it to input $\mathcal P_n(\I)$ by letting
  process $p$ starting with $v_i$ follow the instructions of process
  $i$ in the given algorithm $Algo$.
\end{proof}

\section{Geometric Definition of Simplicial Complexes}
\subsection{Standard Definitions}
\label{sec:GeometricalDef}
We present here classical definitions of geometric complexes and provide a
link between distributed computability and such geometric setting, as our general computability setting could need infinite
complexes%; such have proved useful in this context.
We fix $N\in\N$ and  denote $\dB(x,r) = \{y \in X
| d(x,y) \le r\}$ with $x \in \R^N, r\in\R$ and $d(x,y)$ the Euclidean
distance on $\R^N$.

\begin{definition}[Geometric Simplex]
 Let $n\in\N$.
 A finite set $\sigma=\{x_0,\dots,x_n\}\subset\R^N$ is called a \emph{simplex} of dimension $n$ if the vectors
 $\{x_1-x_0,\dots,x_n-x_0\}$ are linearly independent.
\end{definition}

Let $|\splx1|$ be the convex hull of $\splx1$ and $Int(\splx1)$
is the interior of $|\splx1|$. The \emph{open star} of $\splx1 \in \cplx_1 : 
$ $St^{\circ}(\splx1,\cplx_1) = \bigcup_{\splx2 \in \cplx_1, 
\splx1 \subseteq \splx2} Int(\splx2)$. 
Let $\mathbb{S}^n$ be ``the'' simplex of dimension $n$~: through
this paper we assume a fixed embedding in $\R^N$ for 
$\mathbb{S}^n=(x_0^*,\dots,x_n^*)$ and a diameter $diam(\mathbb{S}^n)$ at 1.

\begin{definition}[\cite{Munkres84}]
  \label{def:simcomplex}
A \emph{simplicial complex} is a collection $C$ of \emph{simplices}
such that~:
\begin{enumerate}[(a)]
  \item If $\sigma\in C$ and $\sigma'\subset\sigma$, then $\sigma'\in C$,
  \item If $\sigma,\tau\in C$ and $|\sigma|\cap|\tau|\neq\emptyset$ then there exists $\sigma'\in C$ such that
   \begin{itemize}
    \item
      $|\sigma|\cap|\tau|=|\sigma'|$,
    \item
      $\sigma'\subset\sigma, \sigma'\subset\tau.$
   \end{itemize}
\end{enumerate}
\end{definition}

For any simplicial complex $C$, we can associate a set of geometric
points $geo(C) = \bigcup_{\splx1 \in C} |\splx1|$.
We will use $geo(C)$ or $|C|$ interchangeably for finite complexes\footnote{As a set of points $geo(C)$ corresponds to the points of the topological space $|C|$ that is classically called the \emph{geometric realization}.
When $C$ is finite, $geo(C)$ has the same topology (in $\R^N$) as the geometric
realization $|C|$. It could not be the case if $C$ is infinite, see
\cite{CG-24}. 
}.
Let $A$, $B$ be  finite simplicial complexes and $\delta$ a simplicial map from $A$ to $B$. It 
can be extended to $f : |A| \rightarrow |B|$ using \emph{barycentric extension}. 
Let $\splx1 = \{x_0, \dots, x_n\}$ a simplex of $A$. Since any 
$y \in |\splx1|$ could be obtained as $y = \sum_{i = 0}^{d} t_i . x_i$ with 
$t_i \in [0,1]$ and $\sum_{i = 0}^{d} t_i = 1$, we set 
$f(y) = \sum_{i = 0}^{d} t_i . \delta(x_i)$. 
Let $X\subset\R^N$, a function $f : X \rightarrow |\cplx_2|$
respects a carrier map $\Delta : \cplx_1 \rightarrow 2^{\cplx_2}$
with $X \subseteq  |\cplx_1|$, if
$\forall \splx1 \in \cplx_1, f(|\splx1| \cap X) \subseteq
\Delta(\splx1)$.  

\begin{definition}[Subdivision]\cite[Def 3.6.1]{HKRbook}
  \label{def:subdiv}
Let $\cplx_1, \cplx_2$ be two geometric simplicial complexes. We say 
that $\cplx_2$ is a subdivision of $\cplx_1$ if : 
%\begin{itemize}
%	\item
$geo(\cplx_1) = geo(\cplx_2)$,
%\item
and
the geometrization of each simplex of $\cplx_1$ is the union of the geometrization of
finitely many simplices of $\cplx_2$.
%\end{itemize}
\end{definition}

In distributed computing there are two subdivisions that are used, the 
barycentric Subdivision and the Standard Chromatic Subdivision. 
In this paper, the barycentric is useful for another reason than being the 
subdivision used to represent colorless computability (see section 
\ref{sec:geomtric_t-res}).
A complete presentation of the second one is in Section
\ref{sect:chr}.

\begin{definition}[Barycentric Subdivision]
  Let $C$ an abstract simplicial complex, its barycentric subdivision 
  $Bary(C)$ is the abstract simplicial complex, whose vertices are the nonempty 
  simplices of $C$. A $(\ell +1)$ tuple $\{\splx1_0, \dots , 
  \splx1_{\ell}\}$ is a simplex of $Bary(C)$ if and only if the tuple 
  can be indexed by containment.
  \end{definition}

%%  Hence, we add an 
%% additional notation for the top-part of the barycentric subdivision : 
%% $\mathcal{U}_t(C)$. Let $V(\mathcal{U}_t(C)) = \{ \splx2 \in C \, | \,
%%  \dim(\splx2) \geq n-t \}$, 
%% and the simplices are sets of vertices that can be ordered by Containment.  

\subsection{Geometric Encoding of Iterated Immediate Snapshots}% Configurations}
\label{sec:geo}

Here, we present the connection between executions of the Iterated Immediate 
Snapshot and simplicial complexes. To achieve this, we employ an algorithm called 
the Chromatic Average algorithm. It accepts an execution $w$ in the $IIS$ 
model and produces a geometrical simplex. This has been introduced in 
\cite{CG-geoconf}, for a detailed presentation refer to~\cite{CG-geoconf,CG-24}.
\SetKwFor{Loop}{Loop}{}{EndLoop}

\begin{algorithm}[ht]
  $x \leftarrow x_i^*\in\R^N$\;
%  $r\leftarrow 0$\;
  \Loop{forever}{
%    $r\leftarrow r+1$\;
    \texttt{SendAll}$((i,x))$\;
    $V\leftarrow$\texttt{Receive()} // set of all received messages including its own\; %
    $d\leftarrow sizeof(V) - 1$ // $i$ received $d+1$ messages
    including its own \; 
    %    $x = \frac{1-\frac{d}{2d+1}}{d+1}x + \sum_{(j,x_j)\in V, j\neq i}\frac{1+\frac{1}{2d+1}}{d+1}x_j$\;
    $x = \frac{1}{2d+1}x + \sum_{(j,x_j)\in V, j\neq i}\frac{2}{2d+1}x_j$\;
  }
  \caption{The Chromatic Average Algorithm for process $i$\label{alg:std-loop}}
\end{algorithm}

A given loop of this algorithm corresponds to one execution of the 
Immediate Snapshot protocol. Geometrically, this 
algorithm associates with every possible view of processes and initial 
configuration $\splx1 \in \I$ in the $ImS$ protocol to a given vertex in 
the Standard Chromatic Subdivision of $\splx1$. We present in more detail this 
subdivision in the Appendix Section \ref{sect:chr}. The equivalence between this 
subdivision and the $ImS$ model can be seen in~\cite{Koz12}.

\begin{figure}[ht]
  \begin{subfigure}{.49\textwidth}
  \begin{tikzpicture}[scale=0.4]
    \path[fill=lightgray,opacity=0.5] (10,0) -- (5.0, 1.7320508075688772) -- (6.6666666,0);

    \node[proc,g,label=north:] (0) at (5.0, 8.660254037844386) {};
    \node[proc,b,label=south:] (1) at (0, 0) {};
    \node[proc,n] (2) at (10, 0) {};

          \node[proc,g,label=left:] (3) at (5.0, 1.7320508075688772) {};
%      \node[proc,b] (4) at (6.0, 3.4641016151377544) {};
%      \node[proc,n] (5) at (4.0, 3.4641016151377544) {};
    %\node[proc,n,label=south:] (6) at (3.3333333,0) {};
          \node[proc,b] (7) at (6.6666666,0) {};

          %      \draw (3) -- (4); % [ultra thick]
%      \draw (4) -- (5); % [ultra thick]
%      \draw (5) -- (3); % [ultra thick]
    %\draw (1) -- (6);
          \draw (1) -- (7); % [ultra thick];
                \draw (7) -- (2);
    \draw (0) -- (2);
    \draw (1) -- (0);
          \draw (3) -- (7);
          \draw (3) -- (2);
                \node[proc,g] (8) at (9.25,8) {}; %(8,666 7.73)
    \node[proc,n,label=south west:$G$] (9) at (10.5,6) {};	%(12,6)
    \node[proc,b] (10) at (8,6) {}; %(7,7.73)
          \draw[rel,thick,->] (9) -- (8);
          \draw[rel,thick,->] (9) -- (10);
                \draw[rel,thick,<-] (8) -- (10);

    \node at (barycentric cs:8=1,9=1,10=1) (C) {};
    \node[label=south:$geo(G)(\mathbb{S}^2)$] at (barycentric cs:3=1,2=2,7=2) (A) {};
    \draw[dashed,<->] (A) -- (C);
  \end{tikzpicture}
\caption{Association between an instant graph $G \in ImS_2$ and a simplex 
 in $\chr(\mathbb{S}^2)$\label{execchr}.}
\end{subfigure}
\hfill
\begin{subfigure}{.49\textwidth}
\begin{tikzpicture}[scale=0.47]
      \node[proc,g,label=north:] (0) at (5.0, 8.660254037844386) {};
      \node[proc,b,label=south:] (1) at (0, 0) {};
      \node[proc,n] (2) at (10, 0) {};
      
      % Middle triangle
      \node[proc,g,label=left:] (3) at (5.0, 1.7320508075688772) {};
      \node[proc,b] (4) at (6.0, 3.4641016151377544) {};
      \node[proc,n] (5) at (4.0, 3.4641016151377544) {};
      
      \draw[rel,thick,<->] (3) -- (4); 
      \draw[rel,thick,<->] (4) -- (5); 
      \draw[rel,thick,<->] (5) -- (3);

      %White and Black line
      \node[proc,n,label=south:] (6) at (3.3333333,0) {}; 
      \node[proc,b] (7) at (6.6666666,0) {};
      
      \draw[rel,thick,->] (1) -- (6); 
      \draw[rel,thick,<->] (6) -- (7); 
      \draw[rel,thick,<-] (7) -- (2); 
      
      % White and Grey line
      
      \node[proc,b] (8) at(3.333,5.773) {};
      \node[proc,g] (9) at(1.66,2.886) {};
      
      \draw[rel,thick,->] (0) -- (8); 
      \draw[rel,thick,<->] (8) -- (9); 
      \draw[rel,thick,<-] (9) -- (1);

      % Black and Grey line  
      
      \node[proc,n] (10) at(6.666,5.773) {};
      \node[proc,g] (11) at(8.3333,2.886) {};
      
      \draw[rel,thick,->] (0) -- (10); 
      \draw[rel,thick,<->] (10) -- (11); 
      \draw[rel,thick,<-] (11) -- (2); 
      
      % Point to center line 
          %Grey
      \draw[rel,thick,->] (0) -- (4); 
      \draw[rel,thick,->] (0) -- (5);
          %White
      \draw[rel,thick,->] (1) -- (5); 
      \draw[rel,thick,->] (1) -- (3); 
          %Black
      \draw[rel,thick,->] (2) -- (3); 
      \draw[rel,thick,->] (2) -- (4); 
      
      % segment to center line 
      
          %Grey
      \draw[rel,thick,<-] (3) -- (6); 
      \draw[rel,thick,<-] (3) -- (7); 
          %White
      \draw[rel,thick,<-] (4) -- (10); 
      \draw[rel,thick,<-] (4) -- (11); 
          %Black
      \draw[rel,thick,<-] (5) -- (8); 
      \draw[rel,thick,<-] (5) -- (9); 

    % Information put in the simplex
    
    \node[font = {\small}] at (4,6) {$G_1$};
    \node[font = {\small}] at (5,6) {$G_2$};
    \node[font = {\small}] at (6,6) {$G_3$};

    \node[font = {\small}] at (8,2.2) {$G_5$};
    \node[font = {\small}] at (7.5,1.5) {$G_6$};
    \node[font = {\small}] at (7.2,0.5) {$G_7$};

    \node[font = {\small}] at (2.8,0.5) {$G_9$};
    \node[font = {\small}] at (2.5,1.5) {$G_{10}$};
    \node[font = {\small}] at (2,2.3) {$G_{11}$};

    \node[font = {\small}] at (3,4) {$G_{12}$};
    \node[font = {\small}] at (7,4) {$G_4$};
    \node[font = {\small}] at (5,0.7) {$G_8$};

    \node[font = {\small}] at (5,3) {$G_{13}$};
            
\end{tikzpicture}
\caption{Standard chromatic subdivision construction for dimension
  $2$ with all corresponding instant graphs\label{allexecchr}.}
\end{subfigure}
\end{figure}

On Fig.~\ref{execchr}, we represent the mapping of one instant graph $G$
from an execution in $IIS$ to a simplex in $\chr(\mathbb{S}^2)$. On Fig.~\ref{allexecchr}, these associations are presented for every possible instant graph of $ImS_2$.
The mapping of a prefix of size $t$ of an execution $w \in IIS_n$ to 
$\splx1 \in \chr^t \mathbb{S}^n$ is called the \emph{geometrization} of 
$w_{|t}$, denoted as $geo(w_{|t})(\mathbb{S}^n)$.

\subsection{Geometrization of Infinite Executions and a Topology for $IIS_n$}
\label{sec:geotopo}

%As stated before, infinite executions can be cumbersome to geometrize, 
Since~\cite{CG-geoconf} the \emph{geometrization} approach has been shown to 
be a fruitful way to handle (the limit of) iterated executions. 
From $geo(w_{|t})(\mathbb{S}^n)$ that work on prefixes of execution, we can 
take the limit on the size of prefixes, $geo(w) = lim_{t \rightarrow \infty} 
geo(w_{|t})(\mathbb{S}^n)$. This operation is well defined, as it makes every 
execution  converge to a geometric point (see~\cite{CG-geoconf} for more 
detail). 

The \emph{geometrization topology} is defined on $IIS_n$ by
considering as open sets the sets $geo^{-1}(\Omega)$ where $\Omega$ is
an open set of $\R^N$.  A collection of sets can define a topology
when any union of sets of the collection is in the collection, and
when any finite intersection of sets of the collection is in the
collection. This is straightforward for a collection of inverse images
of a collection that satisfies these properties.
Note this also makes $geo$ continuous by definition.

The previous construction took $\mathbb{S}^n$ as input, we extend the previous 
definition to any input simplices. For $\ma \subseteq IIS_n$ and 
$\splx1 \in \I$ with $\dim(\splx1) = n$ we can define $\forall w \in \ma, 
geo(w)(\splx1)$ by using a mapping from $\mathbb{S}^n$ to $\splx1$
which maps a vertex $i$ of $\mathbb{S}^n$ to $v_i$ with color $i$ in $V(\splx1)$ (this is 
called the characteristic map of $\splx1$). Hence, $geo(\mathcal{I}
\times \ma)$ is defined as $\bigcup_{w \in \ma, \splx1 \in \mathcal{I}}
\varphi_{\splx1}(geo(w))$. This construction associates to any set 
of executions $\I\times\ma$ a topological subspace of $\R^N$.

%\medskip
Executions in non-compact sub-models of IIS may not terminate at the
same round. The well-known correspondence between terminating
algorithms and complexes could therefore yield infinite complexes.
In this iterated subdivision framework, such (possibly infinite) complexes are called
\emph{terminating subdivisions} and were first introduced in \cite{GKM14}. For this 
article, we use the combinatorial definition of $IIS$-terminating subdivision from~\cite{CG-24}.
Given a complex $C$, let 
$C(T) = \bigcup_{\splx1 \in C, V(\splx1) \subseteq T} \splx1$ with 
$T \subseteq V(C)$ to represent the sub-complex of $C$ formed by the vertices 
in $T$. Let $JOIN(C_1,C_2) = \{ |\splx1 \cup \splx2| | \splx1 \in C_1, 
\splx2 \in C_2\}$.
 We define $EChr$ as the following operator :  
 \begin{equation}
  EChr(T,C) = (\bigcup_{\splx1 \in C} Chr \, \splx1(U)) \cup (
    \bigcup_{\splx1 \in C} JOIN(Chr \, \splx1(U),\splx1(T)) 
  \end{equation}
 Intuitively, the vertices marked as terminated are in $T$. We note 
 $U = V(C) \setminus T$. The operator $EChr$ subdivides with the standard 
 chromatic subdivision the facets that are fully in $U$, does not modify 
 the ones that are fully in $T$ and subdivides in an adequate way the 
 facets containing both.

\begin{definition}[$IIS$-Terminating subdivision~\cite{CG-24}]
  Let $\mathcal{I}$ a simplicial complex.
The sequences $C_0,C_1, \dots$ (collection of simplices) and
$T_0,T_1,\dots $ (collection of increasing set of vertices)
form a \emph{$IIS$-terminating subdivision of $\mathcal{I}$}, if we
have for all $i \in \N$~:
  \begin{enumerate}
	\item $C_0 = \mathcal{I}, T_0 = \emptyset$
	\item $C_{i+1} \subseteq EChr(T_i,C_i)$
	\item $T_{i} \subseteq V(C_{i})$
%	\item $T_i \subseteq T_{i+1}$
  \end{enumerate}
\end{definition}
We say that $\bigcup C_i(T_i)$ is an $IIS$-terminating subdivision complex. 
This is indeed an actual geometric simplicial complex, see \cite{CG-24}.

\section{A General Input-Dependent Colorless Computability Theorem}
\label{sec:extGACT}

In the recent characterization of~\cite{CG-24} of the computability of
colorless tasks, it was presented it in the unique-value
setting.  In the following, we show it is possible to strengthen these
results to the more general input-dependent multi-value setting.

\begin{theorem}\label{th:independantAdv}
  A colorless task $(\I,\mathcal{O},\Delta)$ is solvable on 
  $\mathcal{A} \subseteq \mathcal{P}_n(\I) \times IIS_n$ if and only if there 
  is a continuous function $f : geo(\mathcal{A}) \rightarrow |O|$ that 
  respects $\Delta$.
\end{theorem}

%% Before proving this theorem, let's see what immediate corollary we can deduce :
%% We can extend the result in \cite{CG-24} in the input-dependent adversary

%% \begin{corollary}
%%   A colorless task $(\I,\mathcal{O},\Delta)$ is solvable on 
%%   $\mathcal{A} \subseteq \mathcal{V}_n(\I) \times IIS$ if and only if there 
%%   is a continuous function $f : geo(\mathcal{A}) \rightarrow |O|$ that 
%%   respect $\Delta$.
%% \end{corollary}

%% The unique-value and multi-value presentation have the same solvability if 
%% the adversary satisfies some condition.

%% \begin{corollary}
%%   Let $\ma$ a symmetric adversary then : 
%%   A colorless task $(\I,\Ou,\Delta)$ in a multi-value 
%%   setting solvable on $\ma$ if and only if 
%%   $(\I,\Ou,\Delta)$ is solvable on $\ma$
%% \end{corollary}

We now prove Theorem \ref{th:independantAdv}.  The proofs from
~\cite{CG-24} consider an arbitrary topological space $X$ as input.  Since $\mathcal{A} \subseteq
\mathcal{P}_n(\I) \times IIS_n$ can define a subspace of $\R^N$ by geometrization,
we re-use most of the proof using
$X =  geo(\mathcal{A})$, making some key
adjustments when necessary. In this proof, they extend two concepts
from the proof in the $IIS_n$ model in Chapter 4 of~\cite{HKRbook}.
The first one is an alternative handling of the "continuity" when the output 
space is a simplicial complex. 

\begin{definition}[Star Condition for $\eta$  \cite{CG-24}]
  Let $\eta: X \longrightarrow ]0,+\infty[$ and let $f : X \rightarrow |\mathcal{O}|$, $f$ satisfies the star condition for $\eta$ if  
  $\forall x \in X, \exists v \in V(\mathcal{O}), f(\dB(x,\eta(x)) \cap X) \subseteq St^{\circ}(v).$
\end{definition}

The second one is the concept of ``semi-simplicial approximation''. It is similar to the classical simplicial approximation~\cite{hatcher}, except
here, the input space is not a simplicial complex, but only a
topological space. They use the $\eta$-star condition property to construct 
an $IIS$-terminating subdivision $\mathcal{K}_{\eta}$ that approximate
$X$.

\begin{definition}[semi-simplicial approximation \cite{CG-24}]\label{def:semisimplicialapprox}
  Let $f : X \rightarrow |\mathcal{O}|$ a function.  The function 
  $\psi : V(\mathcal{K}) \rightarrow V(\mathcal{O})$ is a semi-simplicial approximation 
  for $f$ if $\mathcal{K}$ is a
  $IIS$-terminating subdivision that cover $X$, and $\psi$ is a
  simplicial map such that $\forall \splx1 \in \mathcal{K}, f(
  St^{\circ}(\splx1) \cap X) \subseteq St^{\circ}(\psi(\splx1)).$
\end{definition}

We prove that a continuous function $f: geo(\mathcal A)\longrightarrow |\Out|$ implies a distributed algorithm  solving the task,
then, conversely, from a distributed algorithm, we can extract a continuous 
function with Prop.~\ref{prop:algoToContinuous}.

\begin{proposition}[\cite{CG-24}]
  Let $f : geo(\mathcal{A}) \rightarrow |\mathcal{O}|$ a continuous function.
  Then there is $\eta: geo(\mathcal{A}) \longrightarrow ]0,+\infty[$ such that 
  $f$ satisfies the $\eta$-star condition.
\end{proposition}

\begin{proposition}[\cite{CG-24}]
Let $\eta: geo(\mathcal{A}) \longrightarrow ]0,+\infty[$ and let 
$f : geo(\mathcal{A}) \rightarrow |\mathcal{O}|$ a function that satisfies 
the $\eta$-star condition, then $f$ has a semi-simplicial approximation 
$\psi_{\eta} : V(\mathcal{K}_{\eta}) \rightarrow V(\mathcal{O})$.
\end{proposition}

\begin{proposition}[semi-simplicial approximation and carrier map,~\cite{CG-24}]
\label{Approx et carrier}
Let $\eta: geo(\mathcal{A}) \longrightarrow ]0,+\infty[$ and let 
$f: geo(\mathcal{A}) \rightarrow |\mathcal{O}|$ a continuous function that 
respects $\Delta : \mathcal{P}_n(\I) \rightarrow 2^{\mathcal{O}}$ a carrier map. 
Then the semi-simplicial approximation 
$\psi_{\eta} : \mathcal{K}_{\eta} \rightarrow \mathcal{O}$ of $f$ also respects 
$\Delta$.
\end{proposition}

\begin{proposition}
Let $f : geo(\mathcal A) \rightarrow |\mathcal{O}|$ a continuous 
function which respects a carrier $\Delta$ then $(\I,\mathcal{O},\Delta)$ is solvable by an algorithm 
in model $\mathcal A$.
\end{proposition}

These four propositions are already proved in~\cite{CG-24}, hence require no 
proof here. On the other hand, the following proposition needs to be proved again, since 
the setting is more general than the unique-value setting.
We recall that a normalized algorithm is an algorithm where, 
when someone observes a value decided by another process, it immediately decides 
this value. For a colorless task, it does not change the correction of a given 
algorithm.

\begin{proposition}\label{prop:algoToContinuous}
Let $Algo$ a normalized algorithm for solving %the task 
$(\I,\mathcal{O},\Delta)$ against executions $\mathcal{A}\subset \mathcal P_n(\I)\times IIS_n$,
then there exists a continuous function 
 $f: geo(\mathcal{A}) \rightarrow |\mathcal{O}|$ that 
 respects $\Delta$.
\end{proposition}

\begin{proof}
  An algorithm solving a task $(\I,\mathcal{O},\Delta)$ on an adversary 
  $\mathcal{A}$ generates a terminating subdivision
  $\mathcal{K}_{\mathcal{A}}$ and decision function $\varphi : 
  V(\mathcal{K}_{\mathcal{A}}) \rightarrow V(\Ou)$ which is simplicial.
  As emphasized in~\cite{2generals-journal,ACN23,CG-24} we cannot directly 
  take the geometric realization of $\varphi$ to obtain a continuous function 
  if $\mathcal{A}$ is a non-compact message adversary. 
  Consider $v \in V(\mathcal{K}_{\mathcal{A}})$ that decides at round 
  $t$, and $u \in V(St(v,\mathcal{K}_{\mathcal{A}}))$. Using the 
  normalization property : $u$ has decided at round $t+1$ or before. The 
  subcomplex $St(v,\mathcal{K}_{\mathcal{A}})$ is a finite complex, which means 
  that $\varphi(St(v,\mathcal{K}_{\mathcal{A}}))$ we can use the linear extension
  to get a continuous function on $St(v,\mathcal{K}_{\mathcal{A}})$.

  Since the complex is infinite, we cannot directly deduce the
  continuity of the whole function. Consider $x\in geo(\mathcal{A})$,
  and denote $\sigma$ a simplex of $\mathcal{K}_{\mathcal{A}}$ that
  contains $x$. Since $x$ has a neighborhood inside
  $St(v,\mathcal{K}_{\mathcal{A}})$ for some $v$ of $V(\sigma)$, from
  the previous remark, we get that $f$ is continuous at $x$.
  %% Since $\mathcal{K}_{\mathcal{A}}$ cover 
  %% $geo(\mathcal{A})$ we get a continuous function from 
  %% $ geo(\mathcal{A}) \rightarrow |\Ou|$.
\end{proof}

\section{Core-Resilient Adversaries and equivalence of tasks}
\label{sec:coreResMadeSimpler}

\subsection{Geometrical Representation of Core-Resilient Adversary}
\label{sec:geomtric_t-res}

When working with a specific model of computation, a natural question may be, 
"What the geometrization of such model looks like ?" The geometrization of 
$t$-resilient and core-resilient adversaries is not easy to grasp. Hence, one 
goal of this paper is to provide computability equivalence with a model that 
has much nicer geometrization.

Geometrically, we can construct the $t$-resilient adversary by removing all 
simplices of dimension $n-t$ for every round of computation. In other words, 
$geo(R^t_n ) = \bigcap_{i \in\N} |\I| \setminus 
geo(skel^{n-t-1} Chr^i \I)$. 
This yields a fractal-like geometrization of $R^t_n$. 
A detailed presentation of the geometrization of $R_t^n$ is in 
the Appendix, section \ref{sec:Annexe-tRes} for example for $3$ processes and 
one crash. 
For core-resilient adversaries, any set of the core corresponds to a simplex 
$\splx1$ in an input complex $\I$.
Then $\forall \splx2 \subsetneq \splx1$, their images through the iterated subdivisions
are removed from $|\I|$, since it corresponds to a subset of processes that can crash altogether starting from that particular initial configuration. 
As for $R^t_n$ the processes can crash at any moment of the computation; hence, 
simplices in all steps of the geometrization are to be removed,
which also yield a fractal-like space.

These two models have a ``quite complicated'' geometrization space thanks to the "fractal"-part. 
In order to have simpler spaces, we will consider similar models where crashes 
only happened before a given round. 
The $r-$restricted $t$-resilient model is defined as 
$\mathcal{S}^t_r = \{ w \in IIS \, | 
\, K_r(w) \leq t  \}$ with $K_r(w)$ 
the set of processes that are never seen by some process starting from round $r$ excluded, e.g.
$K_0(w)$ is the set  of processes that are crashed at the start of the computation.
We have  simply $geo(\mathcal{S}^t_0) = |\I| \setminus 
|skel^{n-t-1} \I|$.

Similarly, let $\mathcal{H}$ a core-resilient adversary, we denote by 
$\mathcal{H}_r$ the $r-$restricted associated core-adversary, where crashes happen at round $r$. 
Then, $geo(\I \times \mathcal{H}_0) = \I \setminus 
\{ \splx1 \in \I \, | \, |\splx1| \cap geo(\I \times \mathcal{H}) = 
\emptyset\}$. 

Let $\mathcal{H}$ an input-dependent core adversary(\emph{core-dependent} 
in short) on simplicial 
complex $\I$, we denote by $\mathcal C(\mathcal H)$ the 
\emph{condition for $\mathcal H$}, that is 
$\mathcal C(\mathcal H) = \{\sigma\in\In\mid |\sigma|\cap geo(\mathcal H_0) 
\neq \emptyset\}$. By extension of notation, we note 
$\mathcal{U}(\mathcal{H})$ as the part of the barycentric 
subdivision of \I that intersects in $geo(\mathcal{H}_0)$.
More formally~:  
$V(\mathcal{U}(\mathcal{H})) = \{ iso(\splx2)  \, | \,
\splx2\in\mathcal C(\mathcal H)\}$,
where $iso(\splx2)$ is the isobarycenter of simplex $|\splx2|$. 
The simplices of $\mathcal{U}(\mathcal{H})$ corresponds to
sets of isobarycenters of simplices that can be ordered by inclusion.

%% FIXME reprendre si déf au dessus ?
%% Let $\ma \subseteq IIS$, $\I$ a simplicial complex, we define by 
%% $\mathcal{U}(geo(\I \times \ma))$ as the part of the barycentric 
%% subdivision that is present in $geo(\I \times \ma)$. More formally :  
%% $V(\mathcal{U}(geo(\I \times \ma))) = \{ \splx2 \in \I \, | \,
%%  |\splx2| \cap geo(\I \times \ma)  \neq \emptyset\}$, 
%%  and the simplices are sets of vertices that can be ordered by Containement.

\subsection{General Computability Result on core-resilient adversaries}
\label{sec:ComputOn_t-res}

This first theorem shows that models $\mathcal H_0$ associated with core-dependent model $\mathcal H$, ie when crashes happen before
the first round, have the same computability power as $\mathcal U(\mathcal H)$, which corresponds
to a subset of the barycentric subdivision of $\I$.

\begin{theorem}\label{th:equiv_t-res}
  Consider a colorless task $(\I,\Ou,\Delta)$, and $\mathcal H$ a
  core-dependent adversary for $\mathcal P(\I)$.
  The following statements are equivalent
  \begin{enumerate}
    \item $(\I,\Ou,\Delta)$  is solvable on $\mathcal{H}_0$, 
    \item $(\I,\Ou,\Delta)$ is solvable on $geo^{-1}(\mathcal{U}(\mathcal H))$.
  \end{enumerate}
\end{theorem}

\begin{proof}
  Let $Y = geo(\mathcal{H}_0)$. 
  We have that $|\mathcal{U}(\mathcal{H})| \subseteq Y$ 
  which make the $(\Rightarrow)$ direction done.
  
  For the converse, it is sufficient to prove that there is a continuous function 
  $r : Y \rightarrow |\mathcal{U}(\mathcal{H})|$.
  For that, we will construct a collection of functions $r_\sigma$,
  that will project $Y$ on $|\mathcal{U}(\mathcal{H})|$.
  We denote $\mathcal C\subseteq \I$ the condition of $\mathcal H$.
  Let $Q = \{ \splx1 \in \mathcal{C} \, | \, |\splx1| \cap Y = \emptyset\}$. 
  Then for $ \splx1 \in Q,$ let 
  $r_{\splx1} : |St(iso(\splx1), Bary(\mathcal C))| \setminus \{iso(\splx1)\} \rightarrow 
  |Lk(iso(\splx1), Bary(\mathcal C))|$ the projection retract onto the Link from $iso(\splx1)$.
  %  We denote by $v_{\splx1}$ the barycenter of $\splx1$.
 % This is correctly defined since $|St(\splx1, \mathcal{U}(\mathcal{H}))|$ is homeomorphic 
 % to a ball centered on $iso(\splx1)$
 % and, by definition of $Q$, $iso(\splx1) \notin Y$.
  Let $ \tau = (\sigma_0,\sigma_1,\dots,\sigma_d)\in Q\cap Bary(\mathcal C)$.
  Since by definition of the barycentric subdivision, we have
  $\sigma_0\subsetneq\sigma_1\subsetneq\dots\subsetneq\sigma_d\in Q$, we 
  define
  $r_\tau = r_{\sigma_d}\circ\cdots\circ r_{\sigma_0}$.
  This is correctly defined since for $\sigma\in Q,$ $|\sigma|\cap Y = \emptyset$, so $iso(\sigma_{i+1})$ is never in the image of $r_{\sigma_i}$.
  
  We can now make a disjoint sum of the $r_{\tau}$ to construct $r$,
  with all $\tau$ maximal chains in $Bary(\mathcal C)$, by remarking
  that the interiors of $|St(v, Bary(\mathcal C))|,$ with $v\in V(\mathcal C)$
  form a partition of $Y\setminus |\mathcal U(\mathcal H)|$.
\end{proof}

This theorem states that, in a core-adversary model, the crashes that
happened after the first round of communication don't change the
solvability.  The proof is in Appendix~\ref{annex:AddProof}.
\begin{theorem}\label{th:equivCrashStart}
  Let $(\I,\Ou,\Delta)$ a colorless task, $\mathcal H$ a
  core-dependent adversary for $\mathcal P(\I)$.
  The task is solvable on $\mathcal{H}$ if and only if it is solvable on 
  $\mathcal{H}_0$.
\end{theorem}

%\begin{proof}

%\end{proof}

Now we have proved the equivalence of message adversaries $\mathcal H,
\mathcal H_r$ and $\mathcal U(\mathcal H)$,
we give another interpretation on the result from
Th.~\ref{th:equiv_t-res}.

\begin{theorem}
  Consider a colorless task $(\I,\Ou,\Delta)$, and $\mathcal H$ a
  core-dependent adversary for $\mathcal P(\I)$. The task
  $(\I,\Ou,\Delta)$ is solvable on $geo^{-1}(\mathcal U(\mathcal H))$ if and
  only if the task $(\mathcal U(\mathcal H),\Ou, \Delta\circ
  Bary_\I^{-1})$ is solvable against $IIS$.
\end{theorem}

\subsection{Equivalence of distributed tasks}
\label{sec:EquivDistrTask}

Presenting a distributed task as triple $(\I,\Ou,\Delta)$ creates the possibility 
of having tasks that have equivalent computability, which is, they are
solvable on the same message adversaries. In this section, 
we propose multiple propositions/lemmas showing how to transform a task to another one 
that is simpler to analyze but has the same computability.

The carrier map of a task describes where some simplex in the input complex 
can be mapped in the output complex. Then a task is solvable if we are able to 
construct a simplicial map from a terminating subdivision of the input complex to the output 
complex. This first lemma leverages the fact that a simplicial function cannot 
map a simplex to another one of greater dimensions. 

Let $T = (\I,\Ou,\Delta)$ a task is \emph{Non-Expanding}
if $\forall \splx1 \in \I, \dim(\Delta(\splx1)) \leq \dim(\splx1)$. Lemma 
  \ref{lem:dimOfCarrier} implies that any task is equivalent to a 
  non-expanding one. 
Let $(\I,\Ou,\Delta)$ a task, let $\splx1 \in \I$, we set  
$\overline{\Delta}(\splx1) = \{\tau\in\Delta(\splx1) | dim(\tau)\leq dim(\splx1)\}$.

\begin{lemma}\label{lem:dimOfCarrier}
$\forall \mathcal{A} \subseteq \I \times IIS$, the task 
$(\I,\Ou,\Delta)$ is solvable on $\mathcal{A}$ if and only if 
$(\I,\Ou,\overline{\Delta})$ is solvable on $\mathcal{A}$.
\end{lemma}

\begin{proof}
$(\Rightarrow)$ If $(\I,\Ou,\Delta)$ is solvable, then there exists an algorithm 
$Algo$ which yields a terminating subdivision $\mathcal{K}$ and a 
simplicial function $\delta : \mathcal{K} \rightarrow \Ou$. By simpliciality, 
$\forall \splx1 \dim(\delta(\splx1)) \leq \dim(\splx1)$, hence 
$\delta(\splx1) \subseteq \overline{\Delta}(\splx1)$. So $Algo$ also solves $(\I,\Ou,\overline{\Delta})$.

\noindent$(\Leftarrow)$ We have that $\forall \splx1 \in \I, 
\overline{\Delta}(\splx1) \subseteq \Delta(\splx1)$, hence solving
$(\I,\Ou,\overline{\Delta})$ implies solving  $(\I,\Ou,\Delta)$.
\end{proof}

On a distributed task, the carrier map encodes all possible output
values. The next proposition moves this non-determinism on the
vertices to an equivalence of solvability. From $(\I,\Ou,\Delta)$ a
non-expanding task, then $\forall v \in V(\I)$, $\Delta(v)$ is a
collection of vertices from $V(\Out)$. Given a pair $(v,u)$ with $u\in
\Delta(v)$, we set $\Delta_{(v,u)}$ as follows: $\forall \splx1 \in
\I, \splx1 \neq v, \Delta_{(v,u)}(\splx1) = \Delta(\splx1)$ and
$\Delta_{(v,u)}(v) = u$.  Similarly, given a collection $P$ of such
pairs where a vertex of \I appears at most once, we define $\Delta_P$.
We denote $CP$ the set of such collections where all vertices in $\I$ appear.
A task where all input vertices have one possible value by the carrier
map is said to be \emph{vertex-deterministic}. For any collection $P$ in
$CP$, $(\I,\Out,\Delta_P)$ is \emph{vertex-determinism}.

\begin{proposition}\label{prop:break_NonDeterminism}
Let $(\I,\Ou,\Delta)$ a non-expanding task, 
$(\I,\Ou,\Delta)$ is solvable on $\mathcal{A}$ if and only if 
$\exists P \in CP$ such that $(\I,\Ou,\Delta_P)$ is solvable on $\mathcal{A}$.
\end{proposition}

\begin{proof}
  %$(\Rightarrow)$
  Since the task is solvable we have an algorithm that
  yields a terminating subdivision $\mathcal{K}$ and a simplicial map
  $\delta : \mathcal{K} \rightarrow \Ou$ that satisfies $\Delta$. The
  collection $P=\{(v,\delta(v))\}$ satisfies the claim.
%
  %$(\Leftarrow)$
  Conversely, we have
  that $\forall \splx1 \in \I, \Delta_P(\splx1) \subseteq
  \Delta(\splx1)$
\end{proof}

One way to prove that two submodels are equivalent is to construct a continuous 
function between the geometrization of one submodel to the one of the other.
This means finding a continuous  function that preserves the constraints of the carrier map.

\begin{definition}[$\Delta$-compatible function]\label{def:Delta_Compatible}
  Let $(\I,\Ou,\Delta)$ a task, let $Y \subseteq |\I|$ then 
  $f : Y \rightarrow Y$ is \emph{$\Delta$-compatible} if it is continuous 
  and $\forall U \subseteq Y, \Delta(f(U)) \subseteq \Delta(U)$
\end{definition}

\section{Condition Based $k$-set Agreement for core-resilient adversaries}
\label{sec:condBased_kset_tres}
\subsection{Characterization of the solvability}
In this section we look at the solvability of the $k$-set Agreement in 
a condition based context against core-dependent adversaries. This problem was 
introduced in~\cite{condbased} for $t$-resilient models, and the answer can be 
directly inferred for $k +1 \leq t$ from classical result on $t$-resilient 
model~\cite{herlihy_shavit_asynchronous_t-res_1994}. 
For $k +1 > t$, from~\cite{condbased}, we have  some conditions 
that can solve the $k$-set Agreement for $t$-resilient model.
In this section,  we show that deciding if some condition-based
adversary can solve the $k$-set Agreement for a given $t$ is computable.

Let $\mathcal{C}$ a condition for task $(\I,\Ou,\Delta).$
If this task is solvable on $\mathcal{C} \times \ma$ then, $\forall \mathcal{C'} 
\subseteq \mathcal{C}$, this is also solvable.  
Hence, we say that $\mathcal{C}$ is \emph{maximal} for the task 
$T = (\I, \Ou, \Delta)$ if $\mathcal{C}$ solves the task $T$ and 
$\forall \mathcal{C'}, \mathcal{C} \subsetneq \mathcal C'$, $T$ is not
solvable on $\mathcal C'$.
Given a condition $\mathcal C$ on $\mathcal P(\mathbb S^k)$, we
call $\mathcal C$-core-dependent adversary a core-dependent adversary
with support $\mathcal C$. %fixme def support ?

\begin{proposition}\label{prop:KSetIsSimplicialForPsi}
  Let $\mathcal H$ be a $\mathcal C-$core-dependent adversary.
  The $k$-set Agreement task is solvable on condition
  $\mathcal{U}(\mathcal{H})$ against the $IIS$ 
  model if and only if there is a simplicial map 
  $\varphi : \mathcal{U}(\mathcal{H}) \rightarrow
  \partial(\mathbb S^k)$ that respects $\Delta\circ Bary^{-1}$.
\end{proposition}

\begin{proof}
  Using Proposition \ref{prop:break_NonDeterminism}, we get a
  solvable vertex-deterministic task 
  $(\mathcal{U}(\mathcal{H}),\Ou,\Delta_P)$
  with $P \in CP$. Let $\varphi_P$ the simplicial mapping such that
  $\varphi_P(v) = u$ for each pair $(v,u)\in P$.
  By construction of $\Delta_P$, this respects $\Delta\circ Bary^{-1}.$

  The reverse direction is straightforward.
\end{proof}

From Theorem~\ref{th:equiv_t-res} and 
Proposition~\ref{prop:KSetIsSimplicialForPsi}, we can construct an algorithm 
that, from any $\mathcal{C}-$core-dependent adversary $\mathcal H$, tests
if the $k$-set Agreement is solvable by enumerating all $\Delta_P$ and testing
whether $\varphi_P$ is actually simplicial to $\partial(\mathbb S^k)$.

\begin{theorem}
  The problem of deciding if the $k$-set Agreement task is solvable on 
  $\mathcal{H}$ a $\mathcal{C}-$core-dependent adversary 
  is decidable.
\end{theorem}

\subsection{Going back to the 2002 paper}

The~\cite{condbased} paper introduced the condition-based setting against
a $t$-resilient model and asked which conditions solve the $k$-set agreement.

Given a condition $\mathcal{C} \subseteq \mathcal{P}(\I)$, they consider
every set of values of size  at least $n-t+1$
that is present in $\mathcal{C}$ as a vector of size 
$n+1$ with at most $t$ values that are unset.
They consider an associated simplicial complex that
is called $\mathcal{K}in(C,f,k)$
and an example can be found in the Appendix, section \ref{sec:Addfigure}, Figure 
\ref{fig:Sergio-example}.

With this construction, it is proved in \cite{condbased} that if there
is a simplicial function from $\mathcal{K}in(C,f,k)$ to $skel^k \I$,
then a condition $\mathcal{C}$ can solve the $k$-set Agreement for the
$t$-resilient model.

Looking at $\mathcal{K}in(C,f,k)$, we can see that it is the $k$-skeleton of
$\mathcal{U}(C)$ since the barycentric subdivision corresponds to using inclusion-ordered initial simplexes as new simplices.
Hence, Theorem \ref{th:equiv_t-res} shows that the sufficient condition from \cite{condbased} is actually necessary.

Theorem \ref{th:equiv_t-res} provides several 
improvements~: we have a necessary and sufficient characterization that 
works for a larger class of adversaries, the core-dependent adversaries. 

\medskip

Another contribution of~\cite{condbased} is that they provide two example
conditions that generate a set of input configurations that make the 
$k$-set agreement solvable. 

The first condition ($C_1$) uses an order on the input value.
For any proper simplex $\splx1 \in \mathcal{P}(\I)$,
we write $a_1 \dots a_{\ell}$ the input value arranged from
the biggest to the lowest, $a_1 \geq a_2 \dots \geq a_{\ell}$.
Moreover, $\#a_i$ denotes the number of occurrences of the value $a_i$.
If $\sum_{1}^{k} \#a_i > t$, then $\splx1\in C_1$. 

The second condition ($C_2$) works directly on the number of occurrences for 
each input value in a proper simplex $\splx1 \in \mathcal{P}(\I)$. 
Now $a_1$ denotes the number of occurrences of 
the most occurring value, $a_2$ the number of the
second most occurring value etc...
Then $\splx1\in C_2$ if $\sum_{1}^{k} \#a_i - \#a_{k+1} * k > t$.

%% For any input configuration $J$ specified by the condition $C_1$ or 
%% $C_2$ they applied a $t$-resilient adversary to the remaining simplex. 

%% From a set $J$ of proper simplices that solve the $k$-set agreement on a 
%% $t$-resilient model, we can ask if such a set is maximal. A set $J$ is 
%% maximal if $\forall \splx1$ proper simplex of $\I_p$ such that 
%% $\splx1 \notin J$, then $J \cup \splx1$ cannot solve the $k$-set agreement. 

\begin{proposition}
  The condition $C_1$ is a condition for which $k-$set agreement is solvable
  and it is a maximal condition.
\end{proposition}

\begin{proof}
  The condition part is from \cite{condbased},
  the maximality can be seen from the complex $\mathcal{U}(C_1)$. 
  Every set of $n-t+1$ processes will select one value, since this corresponds 
  to one vertex using the Lemma~\ref{prop:KSetIsSimplicialForPsi}. 
  In~\cite{condbased}, the maximum of this set is the 
  "selection" function. 
  We can see that the condition $C_1$
  will remove all proper simplices,
  where there are more than $k$ values that were selected. 
  If we try to remove one less proper simplex $\splx1 \in \mathcal{P}(\I)$, 
  we will have a full simplex in 
  $\mathcal{U}(C) \cup \splx1$ 
  with more than $k$ initial values.
  By the usual set-Agreement impossibility,
  $k-$set agreement is not solvable on  $\mathcal{U}(C) \cup \splx1$. 
\end{proof}

\begin{remark}\label{rmk:ElectForEverySplx}
  We could generalize the condition $C_1$ by choosing a different selection 
  for each set of sizes $n-t+1$ and removing a proper 
  simplex where all of these elections add up to at least $k+1$ values. 
\end{remark}

\begin{proposition}
  Condition $C_2$ is not maximal.
  \end{proposition}
  
  \begin{proof}
    We provide a counter-example, for $3$ processes, $t=1$ and $k=1$ (Consensus task).
    Condition $C_2$ keeps only the simplices
    where all the processes have the same input value.
    Using condition $C_1,$ we know that 
    Consensus is solvable by adding more simplices.
    For example, we can add simplex $(1,1,0)$ and still solve consensus : 
    The simplices with $3$ times the same input decide their initial value and 
    simplex $(1,1,0)$ decides $1$.
  \end{proof}

\section{Conclusion}

In this article, we build on the work of~\cite{CG-24} to extend their 
General Colorless Computability Theorem to an even more general setting. 
Although the proof of this extension uses no new mathematical idea, 
presenting the different settings for colorless distributed computing in 
the topological framework was the main difficulty.
A colorless tasks can be solved with a colored adversary
(or input dependent adversary).
We think that this diversity/complexity is a great testimony of the 
robustness and strength of the topological framework to model distributed 
systems.  

On $t$-resilient models, the reduction to $\mathcal{U}(\mathcal C)$ can be 
interpreted as very close to the classic algorithm 
that is known on $t$-resilient models.
A group of processes of size $n+1-t$ will wait for each other and behave
as one process, hence simulating a system with $t+1$ processes.
Moreover, it seems the approach of~\cite{CG-24} 
using fiber bundles may not be easily extended to the input-dependent 
setting or the core-resilient setting. 
The multiple results of computability equivalence presented in Section  
\ref{sec:EquivDistrTask} may be 
useful tools for future papers using topological methods. The 
$\Delta$-compatible functions are relevant
to produce reductions between different 
adversaries on the same input complex.
Non-expanding tasks and vertex-deterministic tasks allow us to simplify the description of a task and provide direct 
proof (as for Proposition \ref{prop:KSetIsSimplicialForPsi}) 
and, maybe a better intuition of what a task is really about.

The fact that the solvability of $k$-set Agreement on a core-resilient 
model for a particular condition $\mathcal{C}$ is computable leads to 
an interesting question. The proposed algorithm is a brute force method for finding 
all possible simplicial maps. Maybe there is a more efficient algorithm that 
uses the structure provided by the carrier map and the combinatorial 
structure of complexes.
A second complexity question arises~: looking at the existence of a simplicial 
map from $\mathcal{U}(\mathcal{H})$ to $\Ou$ for 
$k$-set Agreement, can we find a class of tasks that has the 
same property ? This research direction may lead to 
tight round-complexity for distributed tasks. 

\paragraph*{Acknowledgements}
The authors wish to sincerely thank Sergio Rajsbaum for suggesting 
condition-based computability as an application of \cite{CG-24}.
This work was partially supported by DUCAT ( Distributed Network Computing 
through the Lens of Combinatorial Topology -- France ANR-20-CE48-0006 )

\bibliography{algoDist.bib}

%\newpage
\appendix

\section{The Standard Chromatic Subdivision} \label{sect:chr}

Here we present the standard chromatic subdivision,~\cite{HKRbook} and
~\cite{Kozlovbook}, as a geometric complex. We start with chromatic subdivisions.

\begin{definition}[Chromatic Subdivision] \label{def:Chrom_Subdv}
  Given $(S,\mathcal P)$ a chromatic simplex, a chromatic subdivision of
  $S$ is a chromatic simplicial complex $(C,\mathcal P_C)$ such that
\begin{itemize}
\item $C$ is a subdivision of $S$ (\ie $|C|=|S|$),
\item $\forall x\in V(S), \mathcal P_C(x)=\mathcal P(x).$
\end{itemize}
\end{definition}

Note that it is not necessary to assume $V(S)\subset V(C)$ here, since
the vertices of the simplex $S$ being extremal points, they are
necessarily in $V(C)$.

We start by defining some geometric transformations of simplices (here
seen as sets of points). The choice of the coefficients will be justified later.
\begin{definition}\label{def:geo}
  Consider a simplex $V=(y_0,\dots,y_d)$ of size $d+1$ in $\R^N$. We
  define the function $\zeta_V:V\longrightarrow
  \mathcal \R^N$ by, for all $i\in[0,d]$
%\begin{align*}
% $$ \zeta_V(y_j) = \frac{1-\frac{d}{2d+1}}{d+1}y_j + \sum_{i\neq j}\frac{1+\frac{1}{2d+1}}{d+1}y_i $$
 $$ \zeta_V(y_i) = \frac{1}{2d+1}y_i + \sum_{j\neq i}\frac{2}{2d+1}y_j $$
%  \zeta_\sigma(p_1) &= \frac{1-\frac{d}{2d+1}}{d+1}p_1 + \sum_{i\neq 1}\frac{1+\frac{1}{2d+1}}{d+1}p_i 
%  \cdots & 
%  \zeta_\sigma(p_{d}) &= \frac{1-\frac{d}{2d+1}}{d+1}p_{d} + \sum_{i\neq d}\frac{1+\frac{1}{2d+1}}{d+1}p_i
%\end{align*}

%% This defines a function $\zeta$ that maps any set $V=\{y_0,\dots,y_d\}$ to the set of same size
%% $\zeta(V)=(\zeta_V(y_0),\dots,\zeta_V(y_d))$.

%We denote $\zeta(S)$ the set $\zeta_S(S)$.

%% Given a subsimplex $T\subset S$ of dimension $d$, with
%% $T=(p_{i_0},p_{i_1}\dots,p_{i_d})$. We denote $\zeta_S(T)$ to be the
%% simplex $(\zeta_S(p_{i_0}),\zeta_S(p_{i_1})\dots,\zeta_S(p_{i_d}))$. 

\end{definition}

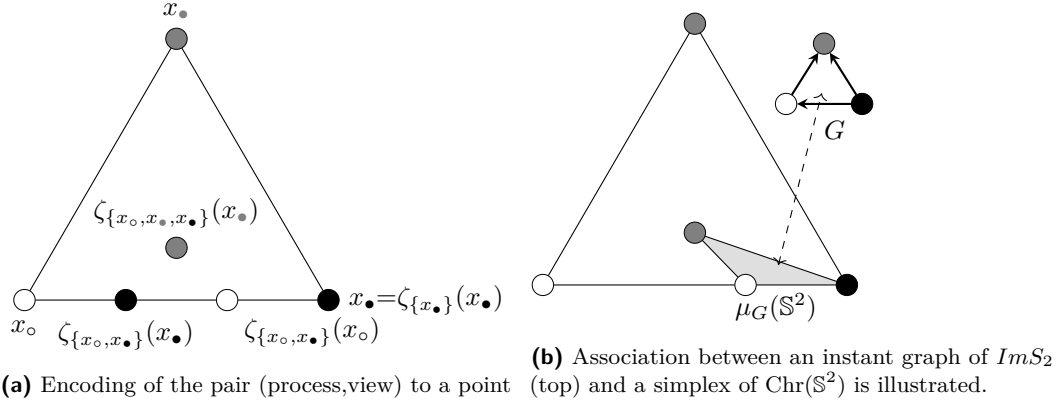
\begin{figure}
	\begin{subfigure}{.49\textwidth}
	\begin{tikzpicture}[scale=.4]
 %\path[fill=lightgray,opacity=0.5] (5.0, 1.7320508075688772) -- (6.0, 3.4641016151377544) -- (4.0, 3.4641016151377544);

    \node[proc,g,label=north:$x_\g$] (0) at (5.0, 8.660254037844386) {};
      \node[proc,b,label=south:$x_\b$] (1) at (0, 0) {};
      \node[proc,n,label=east:$x_\n {=} \zeta_{\{x_\n\}}(x_\n)$] (2) at (10, 0) {};
%      \node[proc,n,label=right:$x_\n=\zeta_{\{x_\n\}}(x_\n)$] (2) at (10, 0) {};
      \node[proc,g,label=north:$\zeta_{\{x_\b,x_\g,x_\n\}}(x_\g)$] (3) at (5.0, 1.7320508075688772) {};
%      \node[proc,b] (4) at (6.0, 3.4641016151377544) {};
%      \node[proc,n] (5) at (4.0, 3.4641016151377544) {};
      \node[proc,n,label=south:$\zeta_{\{x_\b,x_\n\}}(x_\n)$] (6) at (3.3333333,0) {};
      \node[proc,b,label=south east:$\zeta_{\{x_\b,x_\n\}}(x_\b)$] (7) at
      (6.6666666,0) {};
      \node (11) at (8,0.15) {};
      %% \draw (3) -- (4); % [ultra thick]
      %% \draw (4) -- (5); % [ultra thick]
      %% \draw (5) -- (3); % [ultra thick]
      \draw (1) -- (6);
      \draw (6) -- (7); % [ultra thick];
      \draw (7) -- (2);
      \draw (0) -- (2);
      \draw (1) -- (0);
%      \draw (3) -- (7);
%      \draw (3) -- (2);

      %% \node[proc,b] (8) at (8.66666666,6) {};
      %% \node[proc,n] (9) at (12,6) {};
      %% \node[proc,g] (10) at (7,7.7320508075688772) {};
      %% \node (12) at (10,6.6) {};
      %% \draw[rel,thick,->] (9) -- (8);
      %% \draw[rel,thick,->] (9) -- (10);
      %% \draw[rel,thick,->] (8) -- (10);
      %% \draw[dashed,<->] (11) -- (12);
      %      \node at (7.0, 6.0) {$\sigma$};
%      \node at (5.0, 2.9) {};
%      \node[above=3ex] at (3) {$\mu(\sigma)$};
%      \draw[decorate,decoration={brace,amplitude=10pt,raise=4pt}] (2.east) --
%      (1.west) node[below=3ex,midway] {$\dl(\g)$};
%      \draw[decorate,decoration={brace,amplitude=5pt,raise=4pt}] (5.west) --
%      (4.east) node[above=2ex,midway] {$\zeta_{\{\b,\g,\n\}}(\{\b,\g,\n\})$};

  \end{tikzpicture}
	\caption{\label{fig:chr_zeta}Encoding of the pair (process,view) to a point}	
	\end{subfigure}
    \begin{subfigure}{.49\textwidth}
    \begin{tikzpicture}[scale=0.4]
      \path[fill=lightgray,opacity=0.5] (10,0) -- (5.0, 1.7320508075688772) -- (6.6666666,0);

      \node[proc,g,label=north:] (0) at (5.0, 8.660254037844386) {};
      \node[proc,b,label=south:] (1) at (0, 0) {};
      \node[proc,n] (2) at (10, 0) {};

            \node[proc,g,label=left:] (3) at (5.0, 1.7320508075688772) {};
%      \node[proc,b] (4) at (6.0, 3.4641016151377544) {};
%      \node[proc,n] (5) at (4.0, 3.4641016151377544) {};
      %\node[proc,n,label=south:] (6) at (3.3333333,0) {};
            \node[proc,b] (7) at (6.6666666,0) {};

            %      \draw (3) -- (4); % [ultra thick]
%      \draw (4) -- (5); % [ultra thick]
%      \draw (5) -- (3); % [ultra thick]
      %\draw (1) -- (6);
            \draw (1) -- (7); % [ultra thick];
                  \draw (7) -- (2);
      \draw (0) -- (2);
      \draw (1) -- (0);
            \draw (3) -- (7);
            \draw (3) -- (2);
                  \node[proc,g] (8) at (9.25,8) {}; %(8,666 7.73)
      \node[proc,n,label=south west:$G$] (9) at (10.5,6) {};	%(12,6)
      \node[proc,b] (10) at (8,6) {}; %(7,7.73)
            \draw[rel,thick,->] (9) -- (8);
            \draw[rel,thick,->] (9) -- (10);
                  \draw[rel,thick,<-] (8) -- (10);

      \node at (barycentric cs:8=1,9=1,10=1) (C) {};
      \node[label=south:$\mu_G(\mathbb{S}^2)$] at (barycentric cs:3=1,2=2,7=2) (A) {};
      \draw[dashed,<->] (A) -- (C);
    \end{tikzpicture}
\caption{Association between an instant graph of
  $ImS_2$ (top) and a simplex of $\chr(\mathbb{S}^2)$ is illustrated.}
\label{fig:sub_graph3}
  \end{subfigure}
\caption{Construction of $Chr(\mathbb S^2)$ as a geometric encoding for $IIS_2.$\label{fig:sub_chr_2et3d}}
\end{figure}

We now define directly in a geometric way the \emph{standard chromatic
  subdivision} of simplex $S$, where
$S=(x_0,x_1,\dots,x_n)$ . 

The chromatic subdivision $\chr(S)$ for the chromatic simplex
$S=(x_0,\dots,x_n)$ is a simplicial complex defined by the set of
vertices $V(\chr(S)) = 
\{\zeta_{V}(x_i) \mid  i\in [0,n], V\subset V(S), x_i\in V\}.$
The simplices of $\chr(S)$  are the set of $d+1$ points 
$\{\zeta_{V_0}(x_{i_0}),\cdots,\zeta_{V_d}(x_{i_d})\}$ that can be ordered by 
containment.

In Fig.~\ref{fig:sub_chr_2et3d}, we present the construction for
$\chr(\mathbb{S}^2)$. 
For convenience, we associate $\b,\g,\n$ to the processes $0,1,2$
respectively. In Fig.~\ref{fig:chr_zeta}{}, we consider the triangle $x_\b, x_\g, x_\n$ in $\R^2$,
with $x_\b=(0,0)$, $x_\n=(1,0),$ $x_\g=(\frac{1}{2},\frac{\sqrt{3}}{2})$. We have that
$\zeta_{\{x_\b,x_\n\}}(x_\n)=(\frac{1}{3},0)$,
$\zeta_{\{x_\b,x_\n\}}(x_\b)=(\frac{2}{3},0)$
and $\zeta_{\{x_\b,x_\g,x_\n\}}(x_\g) = (\frac{1}{2},\frac{\sqrt{3}}{10})$.
The relation between instant graph $G$ (top) and simplex
$\left\{(\frac{2}{3},0),(1,0),(\frac{1}{2},\frac{\sqrt{3}}{10})\right\}$ (gray area in Fig.~\ref{fig:sub_graph3}) is
detailed in the section \ref{sec:geo}.

In the following, we will be interested in iterations of
$Chr(\mathbb{S}^n)$.

In~\cite{Koz12}, Kozlov showed how the standard chromatic subdivision complex 
relates to Schlegel diagrams (special projections of cross-polytopes), 
and used this relation to prove the standard chromatic subdivision was 
actually a subdivision.
In~\cite[section 3.6.3]{HKRbook}, a general embedding in $\R^{n}$
parameterized by $\epsilon\in\R$ is given for the standard chromatic
subdivision. The geometrization here is done choosing $\epsilon =
\frac{d}{2d+1}$ in order to have ``well balanced'' drawings.

\subsection{Additional figure}
\label{sec:Addfigure}

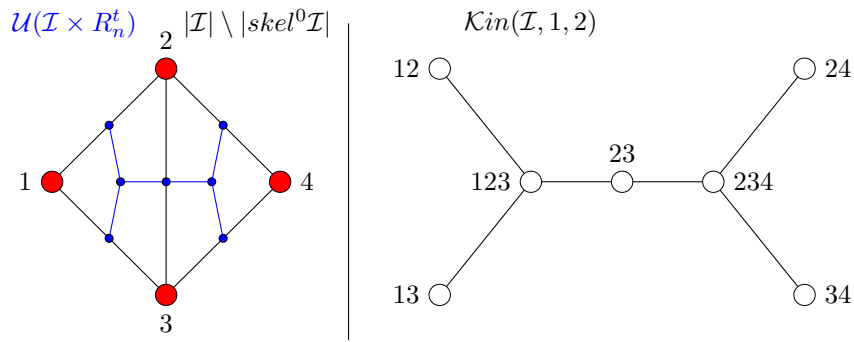
\begin{figure}[ht]
\centering
\begin{tikzpicture}[scale=.6]
    %%%%%%%%%%%%%%%%%%%%%%%%%% I
    \node[] at (2,6) {$|\I| \setminus |skel^{0} \I|$};
    \node[blue] at (-2,6) {$\mathcal{U}(\I \times R^t_n)$};

    \node[proc,b,fill=red,label=left:$1$] (0) at (-2.5,2.5) {};
    \node[proc,b,fill=red,label=north:$2$] (1) at (0,5) {};
    \node[proc,b,fill=red,label=south:$3$] (2) at (0,0) {};
    \node[proc,b,fill=red,label=right:$4$] (3) at (2.5,2.5) {};

    \draw (0) -- (1);
    \draw (1) -- (2);
    \draw (2) -- (0);
    \draw (2) -- (3);
    \draw (1) -- (3);

    %%%%%%%%%%% Barycentric part 

    \node[proc2,fill=blue,label=left:] (4) at (-1.25,3.75) {};
    \node[proc2,fill=blue,label=left:] (5) at (-1.25,1.25) {};
    \node[proc2,fill=blue,label=left:] (6) at (-1,2.5) {};

    \node[proc2,fill=blue,label=north:] (7) at (0,2.5) {};

    \node[proc2,fill=blue,label=right:] (8) at (1,2.5) {};
    \node[proc2,fill=blue,label=right:] (9) at (1.25,1.25) {};
    \node[proc2,fill=blue,label=right:] (10) at (1.25,3.75) {};

    \draw[blue] (4) -- (6);
    \draw[blue] (5) -- (6);
    \draw[blue] (6) -- (7);
    \draw[blue] (7) -- (8);
    \draw[blue] (9) -- (8);
    \draw[blue] (10) -- (8);

    %%%%%%%%%%%%%%%%%%%%%%%%%%%%%%% Separator

    \draw (4,-1) -- (4,6.0);

    %%%%%%%%%%%%%%%%%%%%%%%%%%%% Psi 
    \node[] at (8,6) {$\mathcal{K}in(\I,1,2)$};

    \node[proc,b,label=left:$12$] (4) at (6,5) {};
    \node[proc,b,label=left:$13$] (5) at (6,0) {};
    \node[proc,b,label=left:$123$] (6) at (8,2.5) {};

    \node[proc,b,label=north:$23$] (7) at (10,2.5) {};

    \node[proc,b,label=right:$234$] (8) at (12,2.5) {};
    \node[proc,b,label=right:$34$] (9) at (14,0) {};
    \node[proc,b,label=right:$24$] (10) at (14,5) {};

    \draw (4) -- (6);
    \draw (5) -- (6);
    \draw (6) -- (7);
    \draw (7) -- (8);
    \draw (9) -- (8);
    \draw (10) -- (8);
            
  \end{tikzpicture}
  \caption{Example of $\mathcal{K}in(\I,1,2)$ and $\mathcal{U}_1(\I)$}
  \label{fig:Sergio-example}
\end{figure}
  
On the left of Figure \ref{fig:Sergio-example}, we have an input complex 
with $3$ processes and $t = 1$; hence, 
we removed all the vertices of the $\I$ complex. Moreover we represented in 
blue $\mathcal{U}(\I \times R^t_n)$. On the right, we have every 
set of size $2$ or $3$ that can be formed by any proper simplices of $\I$. 
We can remark that in this example $\mathcal{U}_1$ is equal to 
$\mathcal{K}in(\I,1,2)$.

  \subsection{A complete presentation of the geometrization of $t$-resilient 
  model}
\label{sec:Annexe-tRes}

The fair executions on a simplex $\splx1$ of $\dim(\splx1) = n-t-1$ have 
$n-t$ processes that see each other an infinite amount of time (which means
that they have not crashed). By definition, these executions do not belong to 
$R_t^n$, because t + 1 processes will not ultimately participate.

%% \begin{figure}[ht]
  
%%   \caption{The executions of $IIS_2$}
%%   \label{fig:ex_n=2_t=1_1}
%% \end{figure}

For example, with 3 processes and one crash at most, we obtain the complex
in Figure \ref{allexecchr} the set of execution 
$E = (G_1 + G_2 + G_3)^{\omega} + (G_5 + G_6 + G_7)^{\omega} + (G_9 + G_{10} + 
G_{11})^{\omega}$ are removed in $R_1^2$ and $geo(E)$ correspond to the 
3 points that for this triangle, which is the skeleton of dimension $n-t-1 = 0$ 
of $\I$. 

\begin{figure}[ht]
\centering
\begin{tikzpicture}[scale=.6]
%%%%%%%%%%%%%%%%%%%%%%%%%%%%%

\fill[fill=gray!20] (-7, 8.660254037844386) -- (-12, 0) -- (-2, 0);

% Base triangle
\node[proc2,g,red,label=north:] (0) at (-7.0, 8.660254037844386) {};
\node[proc2,b,red,label=south:] (1) at (-12, 0) {};
\node[proc2,n,red] (2) at (-2, 0) {};

% Middle triangle
\node[proc2,g,red,label=left:] (3) at (-7.0, 1.7320508075688772) {};
\node[proc2,b,red] (4) at (-6.0, 3.4641016151377544) {};
\node[proc2,n,red] (5) at (-8.0, 3.4641016151377544) {};

\draw[] (3) -- (4); 
\draw[] (4) -- (5); 
\draw[] (5) -- (3);

%White and Black line
\node[proc2,n,red,label=south:] (6) at (-8.677,0) {}; 
\node[proc2,b,red] (7) at (-5.333,0) {};

\draw[] (1) -- (6); 
\draw[] (6) -- (7); 
\draw[] (7) -- (2); 

% White and Grey line

\node[proc2,b,red] (8) at(-8.667,5.773) {};
\node[proc2,g,red] (9) at(-10.333,2.886) {};

\draw[] (0) -- (8); 
\draw[] (8) -- (9); 
\draw[] (9) -- (1);

% Black and Grey line  

\node[proc2,n,red] (10) at(-5.334,5.773) {};
\node[proc2,g,red] (11) at(-3.667,2.886) {};

\draw[] (0) -- (10); 
\draw[] (10) -- (11); 
\draw[] (11) -- (2); 

% Point to center line 
      %Grey
\draw[] (0) -- (4); 
\draw[] (0) -- (5);
      %White
\draw[] (1) -- (5); 
\draw[] (1) -- (3); 
      %Black
\draw[] (2) -- (3); 
\draw[] (2) -- (4); 

% segment to center line 

      %Grey
\draw[] (3) -- (6); 
\draw[] (3) -- (7); 
      %White
\draw[] (4) -- (10); 
\draw[] (4) -- (11); 
      %Black
\draw[] (5) -- (8); 
\draw[] (5) -- (9); 

\draw (4.999999999999999, 3.1176914536239786) -- (4.8, 2.7712812921102032);
\draw (4.8, 2.7712812921102032) -- (5.2, 2.7712812921102032);
\draw (5.2, 2.7712812921102032) -- (4.999999999999999, 3.1176914536239786);
\draw (5.0, 1.7320508075688772) -- (4.8, 2.7712812921102032);
\draw (5.0, 1.7320508075688772) -- (5.2, 2.7712812921102032);
\draw (6.0, 3.4641016151377544) -- (5.2, 2.7712812921102032);
\draw (6.0, 3.4641016151377544) -- (4.999999999999999, 3.1176914536239786);
\draw (4.0, 3.4641016151377544) -- (4.999999999999999, 3.1176914536239786);
\draw (4.0, 3.4641016151377544) -- (4.8, 2.7712812921102032);
\draw (5.333333333333333, 2.3094010767585025) -- (5.2, 2.7712812921102032);
\draw (5.666666666666667, 2.886751345948128) -- (5.2, 2.7712812921102032);
\draw (5.0, 1.7320508075688772) -- (5.333333333333333, 2.3094010767585025);
\draw (5.333333333333333, 2.3094010767585025) -- (5.666666666666667, 2.886751345948128);
\draw (5.666666666666667, 2.886751345948128) -- (6.0, 3.4641016151377544);
\draw (5.333333333333334, 3.4641016151377544) -- (4.999999999999999, 3.1176914536239786);
\draw (4.666666666666667, 3.4641016151377544) -- (4.999999999999999, 3.1176914536239786);
\draw (6.0, 3.4641016151377544) -- (5.333333333333334, 3.4641016151377544);
\draw (5.333333333333334, 3.4641016151377544) -- (4.666666666666667, 3.4641016151377544);
\draw (4.666666666666667, 3.4641016151377544) -- (4.0, 3.4641016151377544);
\draw (4.333333333333333, 2.886751345948128) -- (4.8, 2.7712812921102032);
\draw (4.666666666666666, 2.3094010767585025) -- (4.8, 2.7712812921102032);
\draw (4.0, 3.4641016151377544) -- (4.333333333333333, 2.886751345948128);
\draw (4.333333333333333, 2.886751345948128) -- (4.666666666666666, 2.3094010767585025);
\draw (4.666666666666666, 2.3094010767585025) -- (5.0, 1.7320508075688772);

% Center top triangle

\draw (4.999999999999999, 4.50333209967908) -- (4.8, 5.5425625842204065);
\draw (4.8, 5.5425625842204065) -- (5.2, 5.5425625842204065);
\draw (5.2, 5.5425625842204065) -- (4.999999999999999, 4.50333209967908);
\draw (5.0, 8.660254037844386) -- (4.8, 5.5425625842204065);
\draw (5.0, 8.660254037844386) -- (5.2, 5.5425625842204065);
\draw (6.0, 3.4641016151377544) -- (5.2, 5.5425625842204065);
\draw (6.0, 3.4641016151377544) -- (4.999999999999999, 4.50333209967908);
\draw (4.0, 3.4641016151377544) -- (4.999999999999999, 4.50333209967908);
\draw (4.0, 3.4641016151377544) -- (4.8, 5.5425625842204065);
\draw (5.333333333333333, 6.928203230275509) -- (5.2, 5.5425625842204065);
\draw (5.666666666666667, 5.196152422706632) -- (5.2, 5.5425625842204065);
\draw (5.0, 8.660254037844386) -- (5.333333333333333, 6.928203230275509);
\draw (5.333333333333333, 6.928203230275509) -- (5.666666666666667, 5.196152422706632);
\draw (5.666666666666667, 5.196152422706632) -- (6.0, 3.4641016151377544);
\draw (5.333333333333334, 3.4641016151377544) -- (4.999999999999999, 4.50333209967908);
\draw (4.666666666666667, 3.4641016151377544) -- (4.999999999999999, 4.50333209967908);
\draw (6.0, 3.4641016151377544) -- (5.333333333333334, 3.4641016151377544);
\draw (5.333333333333334, 3.4641016151377544) -- (4.666666666666667, 3.4641016151377544);
\draw (4.666666666666667, 3.4641016151377544) -- (4.0, 3.4641016151377544);
\draw (4.333333333333333, 5.196152422706632) -- (4.8, 5.5425625842204065);
\draw (4.666666666666666, 6.928203230275509) -- (4.8, 5.5425625842204065);
\draw (4.0, 3.4641016151377544) -- (4.333333333333333, 5.196152422706632);
\draw (4.333333333333333, 5.196152422706632) -- (4.666666666666666, 6.928203230275509);
\draw (4.666666666666666, 6.928203230275509) -- (5.0, 8.660254037844386);

% Center left triangle

\draw (3.6, 2.0784609690826525) -- (2.6, 1.7320508075688772);
\draw (2.6, 1.7320508075688772) -- (2.8, 1.3856406460551016);
\draw (2.8, 1.3856406460551016) -- (3.6, 2.0784609690826525);
\draw (0.0, 0.0) -- (2.6, 1.7320508075688772);
\draw (0.0, 0.0) -- (2.8, 1.3856406460551016);
\draw (5.0, 1.7320508075688772) -- (2.8, 1.3856406460551016);
\draw (5.0, 1.7320508075688772) -- (3.6, 2.0784609690826525);
\draw (4.0, 3.4641016151377544) -- (3.6, 2.0784609690826525);
\draw (4.0, 3.4641016151377544) -- (2.6, 1.7320508075688772);
\draw (1.6666666666666667, 0.5773502691896257) -- (2.8, 1.3856406460551016);
\draw (3.333333333333333, 1.1547005383792512) -- (2.8, 1.3856406460551016);
\draw (0.0, 0.0) -- (1.6666666666666667, 0.5773502691896257);
\draw (1.6666666666666667, 0.5773502691896257) -- (3.333333333333333, 1.1547005383792512);
\draw (3.333333333333333, 1.1547005383792512) -- (5.0, 1.7320508075688772);
\draw (4.666666666666666, 2.3094010767585025) -- (3.6, 2.0784609690826525);
\draw (4.333333333333333, 2.886751345948128) -- (3.6, 2.0784609690826525);
\draw (5.0, 1.7320508075688772) -- (4.666666666666666, 2.3094010767585025);
\draw (4.666666666666666, 2.3094010767585025) -- (4.333333333333333, 2.886751345948128);
\draw (4.333333333333333, 2.886751345948128) -- (4.0, 3.4641016151377544);
\draw (2.6666666666666665, 2.3094010767585025) -- (2.6, 1.7320508075688772);
\draw (1.3333333333333335, 1.1547005383792515) -- (2.6, 1.7320508075688772);
\draw (4.0, 3.4641016151377544) -- (2.6666666666666665, 2.3094010767585025);
\draw (2.6666666666666665, 2.3094010767585025) -- (1.3333333333333335, 1.1547005383792515);
\draw (1.3333333333333335, 1.1547005383792515) -- (0.0, 0.0);

% Center right triangle

\draw (6.4, 2.0784609690826525) -- (7.4, 1.7320508075688772);
\draw (7.4, 1.7320508075688772) -- (7.2, 1.3856406460551016);
\draw (7.2, 1.3856406460551016) -- (6.4, 2.0784609690826525);
\draw (10.0, 0.0) -- (7.4, 1.7320508075688772);
\draw (10.0, 0.0) -- (7.2, 1.3856406460551016);
\draw (5.0, 1.7320508075688772) -- (7.2, 1.3856406460551016);
\draw (5.0, 1.7320508075688772) -- (6.4, 2.0784609690826525);
\draw (6.0, 3.4641016151377544) -- (6.4, 2.0784609690826525);
\draw (6.0, 3.4641016151377544) -- (7.4, 1.7320508075688772);
\draw (8.333333333333332, 0.5773502691896257) -- (7.2, 1.3856406460551016);
\draw (6.666666666666666, 1.1547005383792512) -- (7.2, 1.3856406460551016);
\draw (10.0, 0.0) -- (8.333333333333332, 0.5773502691896257);
\draw (8.333333333333332, 0.5773502691896257) -- (6.666666666666666, 1.1547005383792512);
\draw (6.666666666666666, 1.1547005383792512) -- (5.0, 1.7320508075688772);
\draw (5.333333333333333, 2.3094010767585025) -- (6.4, 2.0784609690826525);
\draw (5.666666666666667, 2.886751345948128) -- (6.4, 2.0784609690826525);
\draw (5.0, 1.7320508075688772) -- (5.333333333333333, 2.3094010767585025);
\draw (5.333333333333333, 2.3094010767585025) -- (5.666666666666667, 2.886751345948128);
\draw (5.666666666666667, 2.886751345948128) -- (6.0, 3.4641016151377544);
\draw (7.333333333333334, 2.3094010767585025) -- (7.4, 1.7320508075688772);
\draw (8.666666666666666, 1.1547005383792515) -- (7.4, 1.7320508075688772);
\draw (6.0, 3.4641016151377544) -- (7.333333333333334, 2.3094010767585025);
\draw (7.333333333333334, 2.3094010767585025) -- (8.666666666666666, 1.1547005383792515);
\draw (8.666666666666666, 1.1547005383792515) -- (10.0, 0.0);

% Top Black White triangle

\draw (3.9333333333333327, 5.427092530382482) -- (4.266666666666667, 6.004442799572107);
\draw (4.266666666666667, 6.004442799572107) -- (4.133333333333333, 6.466323014923809);
\draw (4.133333333333333, 6.466323014923809) -- (3.9333333333333327, 5.427092530382482);
\draw (5.0, 8.660254037844386) -- (4.266666666666667, 6.004442799572107);
\draw (5.0, 8.660254037844386) -- (4.133333333333333, 6.466323014923809);
\draw (3.333333333333333, 5.773502691896257) -- (4.133333333333333, 6.466323014923809);
\draw (3.333333333333333, 5.773502691896257) -- (3.9333333333333327, 5.427092530382482);
\draw (4.0, 3.4641016151377544) -- (3.9333333333333327, 5.427092530382482);
\draw (4.0, 3.4641016151377544) -- (4.266666666666667, 6.004442799572107);
\draw (4.444444444444445, 7.698003589195011) -- (4.133333333333333, 6.466323014923809);
\draw (3.8888888888888884, 6.735753140545635) -- (4.133333333333333, 6.466323014923809);
\draw (5.0, 8.660254037844386) -- (4.444444444444445, 7.698003589195011);
\draw (4.444444444444445, 7.698003589195011) -- (3.8888888888888884, 6.735753140545635);
\draw (3.8888888888888884, 6.735753140545635) -- (3.333333333333333, 5.773502691896257);
\draw (3.555555555555556, 5.003702332976756) -- (3.9333333333333327, 5.427092530382482);
\draw (3.7777777777777772, 4.233901974057256) -- (3.9333333333333327, 5.427092530382482);
\draw (3.333333333333333, 5.773502691896257) -- (3.555555555555556, 5.003702332976756);
\draw (3.555555555555556, 5.003702332976756) -- (3.7777777777777772, 4.233901974057256);
\draw (3.7777777777777772, 4.233901974057256) -- (4.0, 3.4641016151377544);
\draw (4.333333333333333, 5.196152422706632) -- (4.266666666666667, 6.004442799572107);
\draw (4.666666666666666, 6.928203230275509) -- (4.266666666666667, 6.004442799572107);
\draw (4.0, 3.4641016151377544) -- (4.333333333333333, 5.196152422706632);
\draw (4.333333333333333, 5.196152422706632) -- (4.666666666666666, 6.928203230275509);
\draw (4.666666666666666, 6.928203230275509) -- (5.0, 8.660254037844386);

% Center Black White triangle

\draw (3.2666666666666666, 4.2723919920032305) -- (2.9333333333333327, 3.6950417228136048);
\draw (2.9333333333333327, 3.6950417228136048) -- (2.8, 4.156921938165305);
\draw (2.8, 4.156921938165305) -- (3.2666666666666666, 4.2723919920032305);
\draw (1.6666666666666667, 2.886751345948129) -- (2.9333333333333327, 3.6950417228136048);
\draw (1.6666666666666667, 2.886751345948129) -- (2.8, 4.156921938165305);
\draw (3.333333333333333, 5.773502691896257) -- (2.8, 4.156921938165305);
\draw (3.333333333333333, 5.773502691896257) -- (3.2666666666666666, 4.2723919920032305);
\draw (4.0, 3.4641016151377544) -- (3.2666666666666666, 4.2723919920032305);
\draw (4.0, 3.4641016151377544) -- (2.9333333333333327, 3.6950417228136048);
\draw (2.2222222222222223, 3.8490017945975055) -- (2.8, 4.156921938165305);
\draw (2.7777777777777777, 4.811252243246882) -- (2.8, 4.156921938165305);
\draw (1.6666666666666667, 2.886751345948129) -- (2.2222222222222223, 3.8490017945975055);
\draw (2.2222222222222223, 3.8490017945975055) -- (2.7777777777777777, 4.811252243246882);
\draw (2.7777777777777777, 4.811252243246882) -- (3.333333333333333, 5.773502691896257);
\draw (3.555555555555556, 5.003702332976756) -- (3.2666666666666666, 4.2723919920032305);
\draw (3.7777777777777772, 4.233901974057256) -- (3.2666666666666666, 4.2723919920032305);
\draw (3.333333333333333, 5.773502691896257) -- (3.555555555555556, 5.003702332976756);
\draw (3.555555555555556, 5.003702332976756) -- (3.7777777777777772, 4.233901974057256);
\draw (3.7777777777777772, 4.233901974057256) -- (4.0, 3.4641016151377544);
\draw (3.2222222222222223, 3.2716515254078793) -- (2.9333333333333327, 3.6950417228136048);
\draw (2.4444444444444446, 3.079201435678004) -- (2.9333333333333327, 3.6950417228136048);
\draw (4.0, 3.4641016151377544) -- (3.2222222222222223, 3.2716515254078793);
\draw (3.2222222222222223, 3.2716515254078793) -- (2.4444444444444446, 3.079201435678004);
\draw (2.4444444444444446, 3.079201435678004) -- (1.6666666666666667, 2.886751345948129);

% Bottom Black White triangle

\draw (1.9333333333333331, 1.9629909152447271) -- (2.2666666666666666, 2.540341184434353);
\draw (2.2666666666666666, 2.540341184434353) -- (1.4666666666666666, 1.8475208614068024);
\draw (1.4666666666666666, 1.8475208614068024) -- (1.9333333333333331, 1.9629909152447271);
\draw (1.6666666666666667, 2.886751345948129) -- (2.2666666666666666, 2.540341184434353);
\draw (1.6666666666666667, 2.886751345948129) -- (1.4666666666666666, 1.8475208614068024);
\draw (0.0, 0.0) -- (1.4666666666666666, 1.8475208614068024);
\draw (0.0, 0.0) -- (1.9333333333333331, 1.9629909152447271);
\draw (4.0, 3.4641016151377544) -- (1.9333333333333331, 1.9629909152447271);
\draw (4.0, 3.4641016151377544) -- (2.2666666666666666, 2.540341184434353);
\draw (1.1111111111111112, 1.9245008972987527) -- (1.4666666666666666, 1.8475208614068024);
\draw (0.5555555555555556, 0.9622504486493766) -- (1.4666666666666666, 1.8475208614068024);
\draw (1.6666666666666667, 2.886751345948129) -- (1.1111111111111112, 1.9245008972987527);
\draw (1.1111111111111112, 1.9245008972987527) -- (0.5555555555555556, 0.9622504486493766);
\draw (0.5555555555555556, 0.9622504486493766) -- (0.0, 0.0);
\draw (1.3333333333333335, 1.1547005383792515) -- (1.9333333333333331, 1.9629909152447271);
\draw (2.6666666666666665, 2.3094010767585025) -- (1.9333333333333331, 1.9629909152447271);
\draw (0.0, 0.0) -- (1.3333333333333335, 1.1547005383792515);
\draw (1.3333333333333335, 1.1547005383792515) -- (2.6666666666666665, 2.3094010767585025);
\draw (2.6666666666666665, 2.3094010767585025) -- (4.0, 3.4641016151377544);
\draw (3.2222222222222223, 3.2716515254078793) -- (2.2666666666666666, 2.540341184434353);
\draw (2.4444444444444446, 3.079201435678004) -- (2.2666666666666666, 2.540341184434353);
\draw (4.0, 3.4641016151377544) -- (3.2222222222222223, 3.2716515254078793);
\draw (3.2222222222222223, 3.2716515254078793) -- (2.4444444444444446, 3.079201435678004);
\draw (2.4444444444444446, 3.079201435678004) -- (1.6666666666666667, 2.886751345948129);

% Black Gray left triangle

\draw (2.3333333333333335, 0.3464101615137754) -- (3.3333333333333335, 0.6928203230275508);
\draw (3.3333333333333335, 0.6928203230275508) -- (2.666666666666667, 0.6928203230275508);
\draw (2.666666666666667, 0.6928203230275508) -- (2.3333333333333335, 0.3464101615137754);
\draw (5.0, 1.7320508075688772) -- (3.3333333333333335, 0.6928203230275508);
\draw (5.0, 1.7320508075688772) -- (2.666666666666667, 0.6928203230275508);
\draw (0.0, 0.0) -- (2.666666666666667, 0.6928203230275508);
\draw (0.0, 0.0) -- (2.3333333333333335, 0.3464101615137754);
\draw (3.3333333333333335, 0.0) -- (2.3333333333333335, 0.3464101615137754);
\draw (3.3333333333333335, 0.0) -- (3.3333333333333335, 0.6928203230275508);
\draw (3.333333333333333, 1.1547005383792512) -- (2.666666666666667, 0.6928203230275508);
\draw (1.6666666666666667, 0.5773502691896257) -- (2.666666666666667, 0.6928203230275508);
\draw (5.0, 1.7320508075688772) -- (3.333333333333333, 1.1547005383792512);
\draw (3.333333333333333, 1.1547005383792512) -- (1.6666666666666667, 0.5773502691896257);
\draw (1.6666666666666667, 0.5773502691896257) -- (0.0, 0.0);
\draw (1.1111111111111112, 0.0) -- (2.3333333333333335, 0.3464101615137754);
\draw (2.2222222222222223, 0.0) -- (2.3333333333333335, 0.3464101615137754);
\draw (0.0, 0.0) -- (1.1111111111111112, 0.0);
\draw (1.1111111111111112, 0.0) -- (2.2222222222222223, 0.0);
\draw (2.2222222222222223, 0.0) -- (3.3333333333333335, 0.0);
\draw (3.8888888888888884, 0.5773502691896257) -- (3.3333333333333335, 0.6928203230275508);
\draw (4.444444444444445, 1.1547005383792512) -- (3.3333333333333335, 0.6928203230275508);
\draw (3.3333333333333335, 0.0) -- (3.8888888888888884, 0.5773502691896257);
\draw (3.8888888888888884, 0.5773502691896257) -- (4.444444444444445, 1.1547005383792512);
\draw (4.444444444444445, 1.1547005383792512) -- (5.0, 1.7320508075688772);

% Black Grey Middle triangle

\draw (4.999999999999999, 0.3464101615137754) -- (4.666666666666666, 0.6928203230275508);
\draw (4.666666666666666, 0.6928203230275508) -- (5.333333333333332, 0.6928203230275508);
\draw (5.333333333333332, 0.6928203230275508) -- (4.999999999999999, 0.3464101615137754);
\draw (5.0, 1.7320508075688772) -- (4.666666666666666, 0.6928203230275508);
\draw (5.0, 1.7320508075688772) -- (5.333333333333332, 0.6928203230275508);
\draw (6.666666666666666, 0.0) -- (5.333333333333332, 0.6928203230275508);
\draw (6.666666666666666, 0.0) -- (4.999999999999999, 0.3464101615137754);
\draw (3.3333333333333335, 0.0) -- (4.999999999999999, 0.3464101615137754);
\draw (3.3333333333333335, 0.0) -- (4.666666666666666, 0.6928203230275508);
\draw (5.555555555555555, 1.1547005383792512) -- (5.333333333333332, 0.6928203230275508);
\draw (6.11111111111111, 0.5773502691896257) -- (5.333333333333332, 0.6928203230275508);
\draw (5.0, 1.7320508075688772) -- (5.555555555555555, 1.1547005383792512);
\draw (5.555555555555555, 1.1547005383792512) -- (6.11111111111111, 0.5773502691896257);
\draw (6.11111111111111, 0.5773502691896257) -- (6.666666666666666, 0.0);
\draw (5.555555555555555, 0.0) -- (4.999999999999999, 0.3464101615137754);
\draw (4.444444444444445, 0.0) -- (4.999999999999999, 0.3464101615137754);
\draw (6.666666666666666, 0.0) -- (5.555555555555555, 0.0);
\draw (5.555555555555555, 0.0) -- (4.444444444444445, 0.0);
\draw (4.444444444444445, 0.0) -- (3.3333333333333335, 0.0);
\draw (3.8888888888888884, 0.5773502691896257) -- (4.666666666666666, 0.6928203230275508);
\draw (4.444444444444445, 1.1547005383792512) -- (4.666666666666666, 0.6928203230275508);
\draw (3.3333333333333335, 0.0) -- (3.8888888888888884, 0.5773502691896257);
\draw (3.8888888888888884, 0.5773502691896257) -- (4.444444444444445, 1.1547005383792512);
\draw (4.444444444444445, 1.1547005383792512) -- (5.0, 1.7320508075688772);

%  Right Black Grey triangle

\draw (7.666666666666666, 0.3464101615137754) -- (7.333333333333333, 0.6928203230275508);
\draw (7.333333333333333, 0.6928203230275508) -- (6.666666666666666, 0.6928203230275508);
\draw (6.666666666666666, 0.6928203230275508) -- (7.666666666666666, 0.3464101615137754);
\draw (5.0, 1.7320508075688772) -- (7.333333333333333, 0.6928203230275508);
\draw (5.0, 1.7320508075688772) -- (6.666666666666666, 0.6928203230275508);
\draw (6.666666666666666, 0.0) -- (6.666666666666666, 0.6928203230275508);
\draw (6.666666666666666, 0.0) -- (7.666666666666666, 0.3464101615137754);
\draw (10.0, 0.0) -- (7.666666666666666, 0.3464101615137754);
\draw (10.0, 0.0) -- (7.333333333333333, 0.6928203230275508);
\draw (5.555555555555555, 1.1547005383792512) -- (6.666666666666666, 0.6928203230275508);
\draw (6.11111111111111, 0.5773502691896257) -- (6.666666666666666, 0.6928203230275508);
\draw (5.0, 1.7320508075688772) -- (5.555555555555555, 1.1547005383792512);
\draw (5.555555555555555, 1.1547005383792512) -- (6.11111111111111, 0.5773502691896257);
\draw (6.11111111111111, 0.5773502691896257) -- (6.666666666666666, 0.0);
\draw (7.777777777777777, 0.0) -- (7.666666666666666, 0.3464101615137754);
\draw (8.88888888888889, 0.0) -- (7.666666666666666, 0.3464101615137754);
\draw (6.666666666666666, 0.0) -- (7.777777777777777, 0.0);
\draw (7.777777777777777, 0.0) -- (8.88888888888889, 0.0);
\draw (8.88888888888889, 0.0) -- (10.0, 0.0);
\draw (8.333333333333332, 0.5773502691896257) -- (7.333333333333333, 0.6928203230275508);
\draw (6.666666666666666, 1.1547005383792512) -- (7.333333333333333, 0.6928203230275508);
\draw (10.0, 0.0) -- (8.333333333333332, 0.5773502691896257);
\draw (8.333333333333332, 0.5773502691896257) -- (6.666666666666666, 1.1547005383792512);
\draw (6.666666666666666, 1.1547005383792512) -- (5.0, 1.7320508075688772);

% Gray White Bottom triangle

\draw (8.066666666666666, 1.9629909152447271) -- (8.533333333333333, 1.8475208614068024);
\draw (8.533333333333333, 1.8475208614068024) -- (7.7333333333333325, 2.540341184434353);
\draw (7.7333333333333325, 2.540341184434353) -- (8.066666666666666, 1.9629909152447271);
\draw (8.333333333333332, 2.886751345948129) -- (8.533333333333333, 1.8475208614068024);
\draw (8.333333333333332, 2.886751345948129) -- (7.7333333333333325, 2.540341184434353);
\draw (6.0, 3.4641016151377544) -- (7.7333333333333325, 2.540341184434353);
\draw (6.0, 3.4641016151377544) -- (8.066666666666666, 1.9629909152447271);
\draw (10.0, 0.0) -- (8.066666666666666, 1.9629909152447271);
\draw (10.0, 0.0) -- (8.533333333333333, 1.8475208614068024);
\draw (7.555555555555555, 3.079201435678004) -- (7.7333333333333325, 2.540341184434353);
\draw (6.777777777777777, 3.2716515254078793) -- (7.7333333333333325, 2.540341184434353);
\draw (8.333333333333332, 2.886751345948129) -- (7.555555555555555, 3.079201435678004);
\draw (7.555555555555555, 3.079201435678004) -- (6.777777777777777, 3.2716515254078793);
\draw (6.777777777777777, 3.2716515254078793) -- (6.0, 3.4641016151377544);
\draw (7.333333333333334, 2.3094010767585025) -- (8.066666666666666, 1.9629909152447271);
\draw (8.666666666666666, 1.1547005383792515) -- (8.066666666666666, 1.9629909152447271);
\draw (6.0, 3.4641016151377544) -- (7.333333333333334, 2.3094010767585025);
\draw (7.333333333333334, 2.3094010767585025) -- (8.666666666666666, 1.1547005383792515);
\draw (8.666666666666666, 1.1547005383792515) -- (10.0, 0.0);
\draw (9.444444444444443, 0.9622504486493766) -- (8.533333333333333, 1.8475208614068024);
\draw (8.88888888888889, 1.9245008972987527) -- (8.533333333333333, 1.8475208614068024);
\draw (10.0, 0.0) -- (9.444444444444443, 0.9622504486493766);
\draw (9.444444444444443, 0.9622504486493766) -- (8.88888888888889, 1.9245008972987527);
\draw (8.88888888888889, 1.9245008972987527) -- (8.333333333333332, 2.886751345948129);

% Grey White middle triangle

\draw (6.733333333333332, 4.2723919920032305) -- (7.199999999999998, 4.156921938165305);
\draw (7.199999999999998, 4.156921938165305) -- (7.0666666666666655, 3.6950417228136048);
\draw (7.0666666666666655, 3.6950417228136048) -- (6.733333333333332, 4.2723919920032305);
\draw (8.333333333333332, 2.886751345948129) -- (7.199999999999998, 4.156921938165305);
\draw (8.333333333333332, 2.886751345948129) -- (7.0666666666666655, 3.6950417228136048);
\draw (6.0, 3.4641016151377544) -- (7.0666666666666655, 3.6950417228136048);
\draw (6.0, 3.4641016151377544) -- (6.733333333333332, 4.2723919920032305);
\draw (6.666666666666666, 5.773502691896257) -- (6.733333333333332, 4.2723919920032305);
\draw (6.666666666666666, 5.773502691896257) -- (7.199999999999998, 4.156921938165305);
\draw (7.555555555555555, 3.079201435678004) -- (7.0666666666666655, 3.6950417228136048);
\draw (6.777777777777777, 3.2716515254078793) -- (7.0666666666666655, 3.6950417228136048);
\draw (8.333333333333332, 2.886751345948129) -- (7.555555555555555, 3.079201435678004);
\draw (7.555555555555555, 3.079201435678004) -- (6.777777777777777, 3.2716515254078793);
\draw (6.777777777777777, 3.2716515254078793) -- (6.0, 3.4641016151377544);
\draw (6.222222222222222, 4.233901974057256) -- (6.733333333333332, 4.2723919920032305);
\draw (6.444444444444445, 5.003702332976756) -- (6.733333333333332, 4.2723919920032305);
\draw (6.0, 3.4641016151377544) -- (6.222222222222222, 4.233901974057256);
\draw (6.222222222222222, 4.233901974057256) -- (6.444444444444445, 5.003702332976756);
\draw (6.444444444444445, 5.003702332976756) -- (6.666666666666666, 5.773502691896257);
\draw (7.222222222222221, 4.811252243246882) -- (7.199999999999998, 4.156921938165305);
\draw (7.777777777777777, 3.8490017945975055) -- (7.199999999999998, 4.156921938165305);
\draw (6.666666666666666, 5.773502691896257) -- (7.222222222222221, 4.811252243246882);
\draw (7.222222222222221, 4.811252243246882) -- (7.777777777777777, 3.8490017945975055);
\draw (7.777777777777777, 3.8490017945975055) -- (8.333333333333332, 2.886751345948129);

% Gray White top triangle

\draw (6.0666666666666655, 5.427092530382482) -- (5.7333333333333325, 6.004442799572107);
\draw (5.7333333333333325, 6.004442799572107) -- (5.866666666666665, 6.466323014923809);
\draw (5.866666666666665, 6.466323014923809) -- (6.0666666666666655, 5.427092530382482);
\draw (5.0, 8.660254037844386) -- (5.7333333333333325, 6.004442799572107);
\draw (5.0, 8.660254037844386) -- (5.866666666666665, 6.466323014923809);
\draw (6.666666666666666, 5.773502691896257) -- (5.866666666666665, 6.466323014923809);
\draw (6.666666666666666, 5.773502691896257) -- (6.0666666666666655, 5.427092530382482);
\draw (6.0, 3.4641016151377544) -- (6.0666666666666655, 5.427092530382482);
\draw (6.0, 3.4641016151377544) -- (5.7333333333333325, 6.004442799572107);
\draw (5.555555555555555, 7.698003589195011) -- (5.866666666666665, 6.466323014923809);
\draw (6.11111111111111, 6.735753140545635) -- (5.866666666666665, 6.466323014923809);
\draw (5.0, 8.660254037844386) -- (5.555555555555555, 7.698003589195011);
\draw (5.555555555555555, 7.698003589195011) -- (6.11111111111111, 6.735753140545635);
\draw (6.11111111111111, 6.735753140545635) -- (6.666666666666666, 5.773502691896257);
\draw (6.444444444444445, 5.003702332976756) -- (6.0666666666666655, 5.427092530382482);
\draw (6.222222222222222, 4.233901974057256) -- (6.0666666666666655, 5.427092530382482);
\draw (6.666666666666666, 5.773502691896257) -- (6.444444444444445, 5.003702332976756);
\draw (6.444444444444445, 5.003702332976756) -- (6.222222222222222, 4.233901974057256);
\draw (6.222222222222222, 4.233901974057256) -- (6.0, 3.4641016151377544);
\draw (5.666666666666667, 5.196152422706632) -- (5.7333333333333325, 6.004442799572107);
\draw (5.333333333333333, 6.928203230275509) -- (5.7333333333333325, 6.004442799572107);
\draw (6.0, 3.4641016151377544) -- (5.666666666666667, 5.196152422706632);
\draw (5.666666666666667, 5.196152422706632) -- (5.333333333333333, 6.928203230275509);
\draw (5.333333333333333, 6.928203230275509) -- (5.0, 8.660254037844386);

% On affiche les points 

% Level 0 
\node[proc2,fill=red] at (5.0, 8.660254037844386) {};
\node[proc2,fill=red] at (0.0, 0.0) {};
\node[proc2,fill=red] at (10.0, 0.0) {};

% Level 1
\node[proc2,fill=red] at (5.0, 1.7320508075688772) {};
\node[proc2,fill=red] at (6.0, 3.4641016151377544) {};
\node[proc2,fill=red] at (4.0, 3.4641016151377544) {};
\node[proc2,fill=red] at (1.6666666666666667, 2.886751345948129) {};
\node[proc2,fill=red] at (3.333333333333333, 5.773502691896257) {};
\node[proc2,fill=red] at (6.666666666666666, 0.0) {};
\node[proc2,fill=red] at (3.3333333333333335, 0.0) {};
\node[proc2,fill=red] at (6.666666666666666, 5.773502691896257) {};
\node[proc2,fill=red] at (8.333333333333332, 2.886751345948129) {};

% Level 2
  % Center
\node[proc2,fill=red] at (4.999999999999999, 3.1176914536239786) {};
\node[proc2,fill=red] at (4.8, 2.7712812921102032) {};
\node[proc2,fill=red] at (5.2, 2.7712812921102032) {};
\node[proc2,fill=red] at (5.666666666666667, 2.886751345948128) {};
\node[proc2,fill=red] at (5.333333333333333, 2.3094010767585025) {};
\node[proc2,fill=red] at (4.666666666666667, 3.4641016151377544) {};
\node[proc2,fill=red] at (5.333333333333334, 3.4641016151377544) {};
\node[proc2,fill=red] at (4.666666666666666, 2.3094010767585025) {};
\node[proc2,fill=red] at (4.333333333333333, 2.886751345948128) {};

  % Center top

\node[proc2,fill=red] at (4.999999999999999, 4.50333209967908) {};
\node[proc2,fill=red] at (4.8, 5.5425625842204065) {};
\node[proc2,fill=red] at (5.2, 5.5425625842204065) {};
\node[proc2,fill=red] at (5.666666666666667, 5.196152422706632) {};
\node[proc2,fill=red] at (5.333333333333333, 6.928203230275509) {};
\node[proc2,fill=red] at (4.666666666666667, 3.4641016151377544) {};
\node[proc2,fill=red] at (5.333333333333334, 3.4641016151377544) {};
\node[proc2,fill=red] at (4.666666666666666, 6.928203230275509) {};
\node[proc2,fill=red] at (4.333333333333333, 5.196152422706632) {};
  
  %Center left
\node[proc2,fill=red] at (3.6, 2.0784609690826525) {};
\node[proc2,fill=red] at (2.6, 1.7320508075688772) {};
\node[proc2,fill=red] at (2.8, 1.3856406460551016) {};
\node[proc2,fill=red] at (3.333333333333333, 1.1547005383792512) {};
\node[proc2,fill=red] at (1.6666666666666667, 0.5773502691896257) {};
\node[proc2,fill=red] at (4.333333333333333, 2.886751345948128) {};
\node[proc2,fill=red] at (4.666666666666666, 2.3094010767585025) {};
\node[proc2,fill=red] at (1.3333333333333335, 1.1547005383792515) {};
\node[proc2,fill=red] at (2.6666666666666665, 2.3094010767585025) {};
  % Center right

\node[proc2,fill=red] at (6.4, 2.0784609690826525) {};
\node[proc2,fill=red] at (7.4, 1.7320508075688772) {};
\node[proc2,fill=red] at (7.2, 1.3856406460551016) {};
\node[proc2,fill=red] at (6.666666666666666, 1.1547005383792512) {};
\node[proc2,fill=red] at (8.333333333333332, 0.5773502691896257) {};
\node[proc2,fill=red] at (5.666666666666667, 2.886751345948128) {};
\node[proc2,fill=red] at (5.333333333333333, 2.3094010767585025) {};
\node[proc2,fill=red] at (8.666666666666666, 1.1547005383792515) {};
\node[proc2,fill=red] at (7.333333333333334, 2.3094010767585025) {};

  % Black White top

\node[proc2,fill=red] at (3.9333333333333327, 5.427092530382482) {};
\node[proc2,fill=red] at (4.266666666666667, 6.004442799572107) {};
\node[proc2,fill=red] at (4.133333333333333, 6.466323014923809) {};
\node[proc2,fill=red] at (3.8888888888888884, 6.735753140545635) {};
\node[proc2,fill=red] at (4.444444444444445, 7.698003589195011) {};
\node[proc2,fill=red] at (3.7777777777777772, 4.233901974057256) {};
\node[proc2,fill=red] at (3.555555555555556, 5.003702332976756) {};
\node[proc2,fill=red] at (4.666666666666666, 6.928203230275509) {};
\node[proc2,fill=red] at (4.333333333333333, 5.196152422706632) {};

  % Black White center
\node[proc2,fill=red] at (3.2666666666666666, 4.2723919920032305) {};
\node[proc2,fill=red] at (2.9333333333333327, 3.6950417228136048) {};
\node[proc2,fill=red] at (2.8, 4.156921938165305) {};
\node[proc2,fill=red] at (2.7777777777777777, 4.811252243246882) {};
\node[proc2,fill=red] at (2.2222222222222223, 3.8490017945975055) {};
\node[proc2,fill=red] at (3.7777777777777772, 4.233901974057256) {};
\node[proc2,fill=red] at (3.555555555555556, 5.003702332976756) {};
\node[proc2,fill=red] at (2.4444444444444446, 3.079201435678004) {};
\node[proc2,fill=red] at (3.2222222222222223, 3.2716515254078793) {};

  % Black White bottom

\node[proc2,fill=red] at (1.9333333333333331, 1.9629909152447271) {};
\node[proc2,fill=red] at (2.2666666666666666, 2.540341184434353) {};
\node[proc2,fill=red] at (1.4666666666666666, 1.8475208614068024) {};
\node[proc2,fill=red] at (0.5555555555555556, 0.9622504486493766) {};
\node[proc2,fill=red] at (1.1111111111111112, 1.9245008972987527) {};
\node[proc2,fill=red] at (2.6666666666666665, 2.3094010767585025) {};
\node[proc2,fill=red] at (1.3333333333333335, 1.1547005383792515) {};
\node[proc2,fill=red] at (2.4444444444444446, 3.079201435678004) {};
\node[proc2,fill=red] at (3.2222222222222223, 3.2716515254078793) {};

  % Black Gray left
  
\node[proc2,fill=red] at (2.3333333333333335, 0.3464101615137754) {};
\node[proc2,fill=red] at (3.3333333333333335, 0.6928203230275508) {};
\node[proc2,fill=red] at (2.666666666666667, 0.6928203230275508) {};
\node[proc2,fill=red] at (1.6666666666666667, 0.5773502691896257) {};
\node[proc2,fill=red] at (3.333333333333333, 1.1547005383792512) {};
\node[proc2,fill=red] at (2.2222222222222223, 0.0) {};
\node[proc2,fill=red] at (1.1111111111111112, 0.0) {};
\node[proc2,fill=red] at (4.444444444444445, 1.1547005383792512) {};
\node[proc2,fill=red] at (3.8888888888888884, 0.5773502691896257) {};

  % Black Gray center
  
\node[proc2,fill=red] at (4.999999999999999, 0.3464101615137754) {};
\node[proc2,fill=red] at (4.666666666666666, 0.6928203230275508) {};
\node[proc2,fill=red] at (5.333333333333332, 0.6928203230275508) {};
\node[proc2,fill=red] at (6.11111111111111, 0.5773502691896257) {};
\node[proc2,fill=red] at (5.555555555555555, 1.1547005383792512) {};
\node[proc2,fill=red] at (4.444444444444445, 0.0) {};
\node[proc2,fill=red] at (5.555555555555555, 0.0) {};
\node[proc2,fill=red] at (4.444444444444445, 1.1547005383792512) {};
\node[proc2,fill=red] at (3.8888888888888884, 0.5773502691896257) {};

  % Black Gray right

\node[proc2,fill=red] at (7.666666666666666, 0.3464101615137754) {};
\node[proc2,fill=red] at (7.333333333333333, 0.6928203230275508) {};
\node[proc2,fill=red] at (6.666666666666666, 0.6928203230275508) {};
\node[proc2,fill=red] at (6.11111111111111, 0.5773502691896257) {};
\node[proc2,fill=red] at (5.555555555555555, 1.1547005383792512) {};
\node[proc2,fill=red] at (8.88888888888889, 0.0) {};
\node[proc2,fill=red] at (7.777777777777777, 0.0) {};
\node[proc2,fill=red] at (6.666666666666666, 1.1547005383792512) {};
\node[proc2,fill=red] at (8.333333333333332, 0.5773502691896257) {};

  % Gray White bottom

\node[proc2,fill=red] at (8.066666666666666, 1.9629909152447271) {};
\node[proc2,fill=red] at (8.533333333333333, 1.8475208614068024) {};
\node[proc2,fill=red] at (7.7333333333333325, 2.540341184434353) {};
\node[proc2,fill=red] at (6.777777777777777, 3.2716515254078793) {};
\node[proc2,fill=red] at (7.555555555555555, 3.079201435678004) {};
\node[proc2,fill=red] at (8.666666666666666, 1.1547005383792515) {};
\node[proc2,fill=red] at (7.333333333333334, 2.3094010767585025) {};
\node[proc2,fill=red] at (8.88888888888889, 1.9245008972987527) {};
\node[proc2,fill=red] at (9.444444444444443, 0.9622504486493766) {};

  % Gray White center

\node[proc2,fill=red] at (6.733333333333332, 4.2723919920032305) {};
\node[proc2,fill=red] at (7.199999999999998, 4.156921938165305) {};
\node[proc2,fill=red] at (7.0666666666666655, 3.6950417228136048) {};
\node[proc2,fill=red] at (6.777777777777777, 3.2716515254078793) {};
\node[proc2,fill=red] at (7.555555555555555, 3.079201435678004) {};
\node[proc2,fill=red] at (6.444444444444445, 5.003702332976756) {};
\node[proc2,fill=red] at (6.222222222222222, 4.233901974057256) {};
\node[proc2,fill=red] at (7.777777777777777, 3.8490017945975055) {};
\node[proc2,fill=red] at (7.222222222222221, 4.811252243246882) {};

  % Gray White top

\node[proc2,fill=red] at (6.0666666666666655, 5.427092530382482) {};
\node[proc2,fill=red] at (5.7333333333333325, 6.004442799572107) {};
\node[proc2,fill=red] at (5.866666666666665, 6.466323014923809) {};
\node[proc2,fill=red] at (6.11111111111111, 6.735753140545635) {};
\node[proc2,fill=red] at (5.555555555555555, 7.698003589195011) {};
\node[proc2,fill=red] at (6.222222222222222, 4.233901974057256) {};
\node[proc2,fill=red] at (6.444444444444445, 5.003702332976756) {};
\node[proc2,fill=red] at (5.333333333333333, 6.928203230275509) {};
\node[proc2,fill=red] at (5.666666666666667, 5.196152422706632) {};
\end{tikzpicture}
\caption{The geometrization of $R_1^2$}
\label{fig:ex_n=2_t=1_2}
\end{figure}
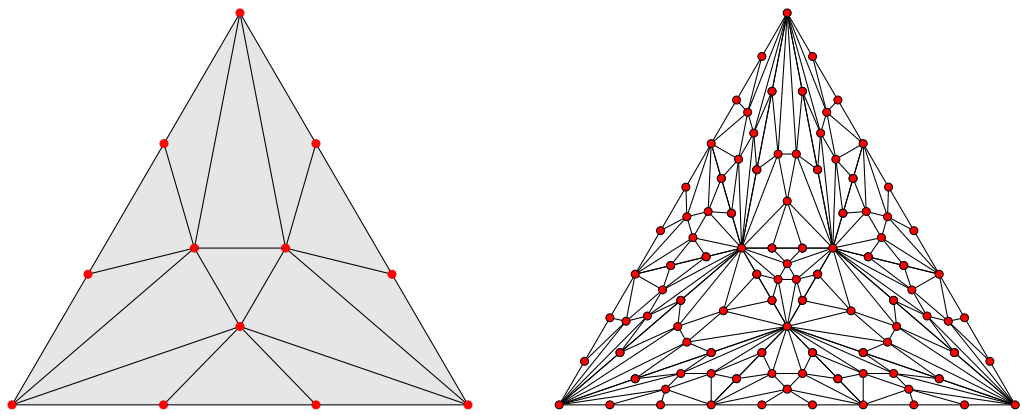     

To obtain the complete geometrization of $R_t^n$ we need to repeat this 
process at every step of the standard chromatic subdivision, for example : 
$G_{13} . E$ is a set of executions that will also be removed (and correspond to 
the 3 vertices of the triangle $G_{13}$). This process yields a fractal like 
structure, another partial representation can be seen in Figure 
\ref{fig:ex_n=2_t=1_2} where all points removed are marked in red for different
levels of subdivision.

\section{Additional Proofs}
\label{annex:AddProof}

\begin{lemma}\label{lem:geometricTool}
  Let $X \subseteq \R^n$, let $\splx2$ a simplex of dimension $t$ such that 
  $|\splx2| \subseteq X$. Let $Z = \{ \splx1_1, \splx1_2, \dots \splx1_{k_1}\}$
  a collection of simplices of $\dim(\splx1_i) \leq n-t-1$ and 
  $\partial(|\splx2|) \subseteq X \setminus Z$.
  Then $\forall k_2 \in \N, k_2 \leq k_1, \exists 
  \varphi_{k_2} : |\splx2| \rightarrow X \setminus \{\splx1_1, \dots 
  \splx1_{k_2}\}$ is a continuous function that is the identity on 
  $\partial(|\splx2|)$
\end{lemma}

The proof of this lemma revolves around taking each intersection of $|\splx2|$
with a $|\splx1_i|$ and constructing a cone to move away from $|\splx1_i|$. 
We then use that lemma to, iteratively, correct any up-part of the chromatic subdivision.
%If the height of each cone gets smaller, you will never cross again a simplex 
%that you already avoided. 

\begin{proof}
  Let $k_2 \in \N$ such that $k_2 \leq k_1$. We will construct a series of 
  function
  $\varphi_{k_2} : |\splx2| \rightarrow  X \setminus \{\splx1_1, \dots 
  \splx1_{k_2}\}$ a continuous property such that $\partial(|\splx2|)$ 
  is the identity.
  We set $\varphi_0(|\splx2|)$ as the identity function. 
  Let $Y_{k_2}$ a connected subspace of $|\splx1_{k_2}| \cap 
  \varphi_{k_2-1}(\splx2)$. We have that $\dim(Y_{k_2}) \leq t-1$ and 
  $\partial(|\splx2|) \cap \splx1_{k_2}$ since $\splx1_{k_2}$ is a simplex. 
  Let $N_{\epsilon}(Y_{k_2}) = \{ x \in X \, | \, 
  MIN d(x,Y_{k_2}) = \epsilon\}$. Then let $\epsilon_0 = + \infty$ and let 
  $\epsilon_i \in \R$ positive, such that 
  $\partial(|\splx2|) \cap N_{\epsilon_{k_2}}(Y_{k_2}) = 
  \emptyset$ and $N_{\epsilon_{k_2}}(Y_{z_2}) \cap \varphi_{k_2-1}(\splx2)$ 
  is a deformation of $\mathbb{S}^{\dim(\partial(\splx2))}$ and 
  $\epsilon_{k_2} < \epsilon_{k_2-1}/3$. Such $\epsilon_{k_2}$ exist since 
  $Y_{k_2}$ is a closed set and $\varphi_{k_2 -1}(|\splx2|) \setminus 
  \partial(|\splx2|)$ is an open set. 
  Now, let $x \in Int(Y_{k_2})$, let $y \in 
  N_{\epsilon}(Y_{k_2}) \setminus \varphi_{k_2-1}(|\splx2|)$. We can construct 
  $B_{k_2}$ : $N_{\epsilon}(Y_{k_2}) \cap \varphi_{k_2-1}(|\splx2|)$ 
  completed in $\varphi_{k_2-1}(|\splx2|)$.
  And $A_{k_2} = N_{\epsilon}(Y_{k_2}) \cap \varphi_{k_2-1}(|\splx2|)$ 
  completed in $ N_{\epsilon}(Y_{k_2})$ and of dimension $t$ going through 
  $y$.
  Then $\varphi_{k_2}(B_{k_2})$ 
  is the projection of $B_{k_2}$ to $A_{k_2}) \setminus B_{k_2}$. 
  This operation is continuous, hence $\varphi_{k_2}(|\splx2|)$ is a continuous 
  function. Moreover, we have that $\forall \splx1_i \in \{ \splx1_1, \dots 
  \splx1_{k_2}\}, \varphi_{k_2}(|\splx2|) \cap |\splx1_i| = \emptyset$ since  
  we have that $\sum_{i+1}^{k_2} \epsilon_j < \epsilon_i$. Hence, 
  $\varphi_{k_2}(|\splx2|)$ is a continuous deformation of $wInt(|\splx2|)$ 
  that does not intersect $ \{ \splx1_1, \dots \splx1_{k_2}\}$.
\end{proof}

\begin{figure}[ht]
\centering
\begin{tikzpicture}[scale=.6]

  \node[blue] at (2.5,4.75) {$\varphi_0(C_1)$};
  \node[blue] at (3.75,1.5) {$C_1$};

 \node[proc2,b] (0) at (0,0) {};
 \node[proc2,b] (1) at (5,0) {};
 \node[proc2,b] (2) at (2.5,4) {};

 \draw (0) -- (1);
 \draw (1) -- (2);
 \draw (2) -- (0);

 \node[proc2,fill, blue] (3) at (2.5, 1.5) {};

 \node[proc2,fill,blue] (4) at (1.25,2) {};
 \node[proc2,fill,blue] (5) at (3.75,2) {};
 \node[proc2,fill,blue] (6) at (2.5,0) {};

 \draw[blue] (4) -- (3);
 \draw[blue] (5) -- (3);
 \draw[blue] (6) -- (3);

 \node[proc2, red, fill, label=south:$\splx1_1$] (7) at (1.875,1.75) {};
 \node[proc2,red,fill,label=right:$\splx1_3$] (8) at (2.1,2) {};
 \node[proc2, red, fill,label=right:$\splx1_2$ ] (9) at (3,0.75) {};

 %%%%%%%%%%%%%%%%
 \node[blue] at (8.5,4.75) {$\varphi_2(C_1)$};
 \node[red] at (11.5,3) {$X \setminus \{\splx1_1,\splx1_2,\splx1_3\}$};
 \node[blue] at (9.75,1.5) {$C_1$};

 \node[proc2,b] (0) at (6,0) {};
 \node[proc2,b] (1) at (11,0) {};
 \node[proc2,b] (2) at (8.5,4) {};

 \draw (0) -- (1);
 \draw (1) -- (2);
 \draw (2) -- (0);

 \node[proc2,fill,blue] (3) at (8.5, 1.5) {};

 \node[proc2,fill,blue] (4) at (7.25,2) {};
 \node[proc2,fill,blue] (5) at (9.75,2) {};
 \node[proc2,fill,blue] (6) at (8.5,0) {};

 \draw[blue] (4) -- (7.6625,1.875) -- (8.05,2.2) -- (8.1875,1.625) -- (3);
 \draw[blue] (5) -- (3);
 \draw[blue] (6) -- (3);

 \node[proc2, red, fill, label=south:$\splx1_1$] (7) at (7.875,1.75) {};
 \node[proc2,red,fill,label=right:$\splx1_3$] (8) at (8.1,2) {};
 \node[proc2, red, fill,label=right:$\splx1_2$ ] (9) at (9,0.75) {};

 %%%%%%%%%%%%%%%%%%

 \node[blue] at (14.5,4.75) {$\varphi_3(C_1)$};
 \node[blue] at (15.75,1.5) {$C_1$};

 \node[proc2,b] (0) at (12,0) {};
 \node[proc2,b] (1) at (17,0) {};
 \node[proc2,b] (2) at (14.5,4) {};

 \draw (0) -- (1);
 \draw (1) -- (2);
 \draw (2) -- (0);

 \node[proc2,fill,blue] (3) at (14.5, 1.5) {};

 \node[proc2,fill,blue] (4) at (13.25,2) {};
 \node[proc2,fill,blue] (5) at (15.75,2) {};
 \node[proc2,fill,blue] (6) at (14.5,0) {};

 \node[] (7) at (13.25,2.5) {};
 \node[] (8) at (13.25,1.5) {};
 \node[] (9) at (13.75,2) {};

 \draw[blue] (4) -- (13.6625,1.875) -- (14.05,2.2) -- (14.06375, 2.1425) --
 (14.35,2.1) -- (14.1325, 1.855)
 
 -- (14.1875,1.625) -- (3);
 \draw[blue] (5) -- (3);
 \draw[blue] (6) -- (3);

 \node[proc2, red, fill, label=south:$\splx1_1$] (7) at (13.875,1.75) {};
 \node[proc2,red,fill,label=right:$\splx1_3$] (8) at (14.1,2) {};
 \node[proc2, red, fill,label=right:$\splx1_2$ ] (9) at (15,0.75) {};

\end{tikzpicture}
\caption{Example of the construction from Lemma \ref{lem:geometricTool}}
\label{fig:ExampleGeometricConstr}
  \end{figure}
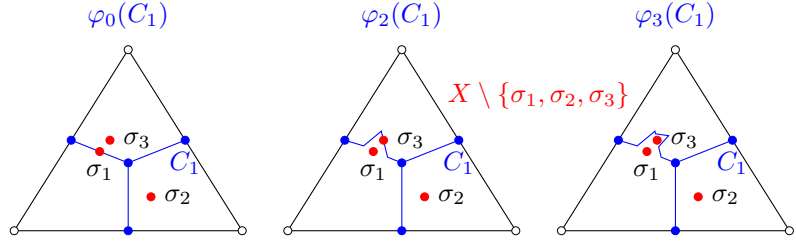

Figure \ref{fig:ExampleGeometricConstr} is a representation of a simple case 
from Lemma \ref{lem:geometricTool}. We have $n = 2,t = 1$, $X$ is the full 
triangle, and we remove $\{\splx1_1, \splx1_2, \splx1_3\}$. On $\varphi_0$ the 
blue segment isn't modified; it gets modified in $\varphi_2$ since $\splx1_1$ 
intersect $\splx2$. The modification is made to form a cone that avoids 
$\splx1_1$, even if it can intersect another point, $\splx1_3$ in this figure. 
Finally, in $\varphi_3, \splx1_3$ is avoided and $\varphi_3(C_1)$ 
doesn't need further modification. 

\medskip
On a core-dependent adversary, we show that restricting crashes happening before the $k-th$ round does not change the computability.
Let $\mathcal{H}_k = \{ w \in IIS \, | \, \forall p \in \Pi \setminus Q(w), 
\forall q \in Q(w), \forall r \in \N, r > k, (p,q) \notin A(w(r))\}$, with 
$Q(w)$ the set of processes that can see each other an infinite amount 
of time in $w$.

\begin{proposition}
  Let $(\I,\Ou,\Delta)$ a colorless task, let $\mathcal C \subseteq \I$ a 
  condition on the input and 
  $\mathcal{H}$ a core-dependent adversary on $\mathcal C$, then, for 
  any $k\in\N$ the task is solvable on $\mathcal{H}_k$ if and only if it is 
  solvable on $\mathcal{H}_0$.
\end{proposition}

\begin{proof}
  We denote by $\mathcal U_k = \mathcal U(Chr^k(\mathcal C))$.
  Note that $|\mathcal U_{k+1}| \subset |\mathcal U_k|$. 
  We prove the result by recursion. \\
  Assume this is true for $k$.  We work simplex by simplex from
  $Chr^{k+1}(\mathcal C)$. Denote $\tau$ the current (maximal) simplex
  and $t$ the size of the current core.  By using
  Lemma~\ref{lem:geometricTool} with $\mathcal U_k$ it is possible to
  derive a new continuous function that avoids all new simplices
  appearing in $Chr^{k+1}(\mathcal C)$. \\
  This way, it is possible to project $|\mathcal U_{k+1}|$ on
  $\mathcal U_0$. So, from Thm~\ref{th:independantAdv}, we get
  equivalent colorless computability on $\mathcal H_{k+1}$ and
  $\mathcal H_0$.
\end{proof}

Using the previous proposition for all $k\in\N$, we have
that this sequence of $\mathcal H_k$ forms a projective limit, a notion from category
theory, where $\mathcal H$ is the limit of $\mathcal H_k$. Since the
projection holds for any $k$, the limit mapping also exists
(topological spaces form a \emph{complete} category) and is also
continuous.

In more detail, we consider the family of inclusion morphisms
$f_{k,k'}:geo(\mathcal H_{k'})\hookrightarrow geo(\times\mathcal
H_{k})$, with $k\leq k'$. Of course, we have $f_{k,k'}\circ f_{k',k''}
= f_{k,k''}$ when $k\leq k'\leq k''$. So this sequence defines a
system of morphisms that is a \emph{diagram} and since the category of
topological spaces with continuous functions is complete~\cite[Section 12.6]{catjoy},
the limit exists and there is a continuous mapping from $\mathcal
U(\mathcal H_0)$ to this limit. The last step is to see that the
limit $\mathcal L$ is actually (homeomorphic to) $geo(\mathcal H)$, since 
$geo(\mathcal H) = \bigcap_k geo(\mathcal H_k)$.

\end{document}